\documentclass[twocolumn]{aastex631}

\usepackage{amsmath,amssymb,bm}

\usepackage{comment}

\usepackage{cancel}
\usepackage{tikz}

\def \be{\begin{equation}}
\def \ee{\end{equation}}
\def \bea{\begin{eqnarray}}
\def \eea{\end{eqnarray}}
\def \la{\langle}
\def \ra{\rangle}

\hypersetup{
    urlcolor=cyan,
    }

\shorttitle{Energy extraction in a thin MAD}
\shortauthors{Dhang, Dexter \& Begelman}

\begin{document}

%\title{The Fate of the Extracted Energy in a Thin Magnetically Arrested Disk}
\title{Energy Extraction from a Black Hole by a Strongly Magnetized Thin Accretion Disk}

\email{prasundhang@gmail.com}
\author[0000-0001-9446-4663]{Prasun Dhang}
\affiliation{JILA, University of Colorado and National Institute of Standards and Technology, 440 UCB, Boulder, CO 80309-0440, USA}
\affiliation{Department of Astrophysical and Planetary Sciences, University of Colorado, 391 UCB, Boulder, CO 80309, USA}

\email{Jason.Dexter@colorado.edu}
\author[0000-0003-3903-0373]{Jason Dexter}
\affiliation{JILA, University of Colorado and National Institute of Standards and Technology, 440 UCB, Boulder, CO 80309-0440, USA}
\affiliation{Department of Astrophysical and Planetary Sciences, University of Colorado, 391 UCB, Boulder, CO 80309, USA}

\email{mitch@jila.colorado.edu}
\author[0000-0003-0936-8488]{Mitchell C. Begelman}
\affiliation{JILA, University of Colorado and National Institute of Standards and Technology, 440 UCB, Boulder, CO 80309-0440, USA}
\affiliation{Department of Astrophysical and Planetary Sciences, University of Colorado, 391 UCB, Boulder, CO 80309, USA}

\begin{abstract}
The presence of a strong, large-scale magnetic field in an accretion flow leads to extraction of the rotational energy of the black hole (BH) through the Blandford-Znajek (BZ) process, believed to power relativistic jets in various astrophysical sources. We study rotational energy extraction from a BH surrounded by a highly magnetized thin disk by performing a set of 3D global GRMHD simulations.  We find that the saturated flux threading the BH has a weaker dependence on BH spin, compared to highly magnetized hot (geometrically thick) accretion flows. Also, we find that only a fraction ($10-70$ per cent) of the extracted BZ power is channeled into the jet, depending on the spin parameter. The remaining energy is potentially used to launch winds or contribute to the radiative output of the disk or corona. Our simulations reveal that the presence of a strong magnetic field enhances the radiative efficiency of the disk, making it more luminous than its weakly magnetized counterpart or the standard disk model. We attribute this excess luminosity primarily to the enhanced magnetic dissipation in the intra-ISCO region.
Our findings have implications for understanding X-ray corona formation and black hole spin measurements, and interpreting black hole transient phenomena.
\end{abstract}


\section{Introduction}
\label{sect:intro}

Accretion onto a centrally gravitating 
object results in the conversion of gravitational potential energy into thermal energy. When the central object is compact, e.g., a black hole (BH) or neutron star, this process yields
%This process is recognized as 
one of the most efficient means of energy generation in the Universe.  
According to the standard model of accretion disks  
(\citealt{Shakura1973, Novikov1973}), 
the radiative efficiency of the disk can be as high as 40\% for a maximally spinning black hole (\citealt{Thorne1974}).
The prevailing consensus among astronomers is that accretion powers a variety of luminous sources in the Universe, including X-ray binaries (XRBs) and Active Galactic Nuclei (AGN). 

In addition to accumulating matter, the accretion process can bring in magnetic field from the surrounding medium (\citealt{Lubow1994, Guilet2012, Cao2011, Dhang2023}) or generate it locally through dynamo action (\citealt{Brandenburg1995, Gressel2015, Dhang2019, Dhang2020, Nathanail2022, Begelman2023, Dhang2024, Jacquemin-Ide2024}) mediated by the magneto-rotational instability (MRI; \citealt{Velikhov1959, Chandrasekhar1960a, Balbus1991}). If a sufficient amount of magnetic flux threads the BH event horizon, the Blandford-Znajek (BZ) mechanism (\citealt{Blandford1977}) extracts energy from the BH's rotational energy, providing an additional means of energy production in accreting sources. Previous studies have shown that the BZ process is particularly efficient for geometrically thick, hot accretion flows (\citealt{Chakrabarti1989, Narayan1994, Blandford_Begelman1999}) saturated with a strong poloidal magnetic flux, known as a magnetically arrested disks (MADs; \citealt{Bisnovatyi-Kogan1974a, Narayan2003, Igumenshchev2003}). MADs around BHs are thought to provide a conducive environment for producing powerful jets in accreting sources (\citealt{Tchekhovskoy2011, McKinney2012}).

The presence of a strong large-scale magnetic flux in the accretion flow not only provides a means of energy extraction from the black hole's rotational energy (\citealt{Blandford1977}) or disk's energy reservoir (\citealt{Blandford1982}), but also tends to alleviate some of the tensions between the standard accretion disk model and observations. Specifically, a large-scale magnetic flux threading the disk leads to efficient angular momentum transport (\citealt{Bai2013}) and vertical support (\citealt{Dexter2019}), resulting in faster inflow and hence shorter evolution times (\citealt{Scepi2024}). Furthermore, the presence of a strong magnetic flux provides stability against thermal instability (\citealt{Begelman_Pringle2007, Sadowski2016}).

Additionally, the standard disk model fails to explain the existence of hard X-ray emission from XRBs and in AGN.
Luminous XRBs and AGN often exhibit multi-color black body emission accompanied by a power law component (\citealt{Remillard2006}, \citealt{Motta2009}). While the optically thick standard disk model can often explain the blackbody part of the spectra (e.g., see \citealt{Hagen2023}), the power-law component necessitates the presence of an optically thin corona . Several studies have proposed that strong magnetic dissipation in the upper disk (\citealt{Miller2000, Jiang2014b, Jiang2019}) or the magnetized plunging region near the innermost stable circular orbit (ISCO) of the black hole (\citealt{Zhu2012, Hankla2022, Mummery2024}) could be responsible for generating the power-law component in the spectra of accreting sources.

Most studies of energy extraction from BH spin
have focused on geometrically thick hot MADs (e.g., \citealt{Tchekhovskoy2011}, \citealt{McKinney2012}, \citealt{White_mad_2019}, \citealt{Narayan2022}), with a few exploring highly magnetized geometrically thin disks using magnetohydrodynamic (MHD; e.g., \citealt{Mishra2020}) and general relativistic magnetohydrodynamic (GRMHD; e.g., \citealt{Avara2016}, \citealt{Liska2022}, \citealt{Scepi2024}) simulations.  While this approach fits with prevailing ideas on low-luminosity XRBs or AGN, where a geometrically thin disk is truncated to a hot accretion flow within some transition radius (\citealt{Nemmen2014}), the geometrically thin disk is expected to extend to the innermost regions in the luminous (high-soft) state of XRBs and AGNs (\citealt{Remillard2006}). In this paper, we present a systematic study of highly magnetized thin disks (thin MADs) around black holes of different spins using a set of GRMHD simulations with an {\em ad hoc} cooling function (\citealt{Noble2009}). The aim of this study is two-fold: (i) to investigate the Blandford-Znajek (BZ) mechanism in a thin MAD, and (ii) to explore  its possible role in powering winds and disk radiation or coronae, in addition to launching jets.
%, and its plausible connection to observations of jets and corona.

We organize the paper as follows. In Section \ref{sect:method}, we discuss the details of the GRMHD simulations employed, the implementation of the cooling function, and the diagnostics used to analyze the simulations. Section \ref{sect:fiducial_run} presents the results for our fiducial run with a black hole spin of $a=0.9$, focusing on characterizing the thin MAD that we achieve after running the simulation using the cooling function for a sufficiently long time. We address the main outcomes of the thin MAD simulations for different black hole spins in Section \ref{sect:results_thin_mad}, primarily discussing the energy extraction from the BH spin and its contribution to launching jets, as well as the highly radiative properties of the thin MAD. In Section \ref{sect:discussion}, we discuss the observational consequences of our study, particularly the plausible manifestation of BZ power in  both launching jets and forming coronae in accreting sources. Finally, we summarize our work in Section \ref{sect:summary}. Additionally, we include appendices that provide further clarification on a few topics not covered in the main text.

\section{Method}
\label{sect:method}

\subsection{Equations solved}
\label{sect:method_eqs}
We solve the ideal GRMHD equations
\bea
\label{eq:mass}
 && \nabla_{\mu} \left(\rho u^{\mu} \right) = 0\\
  \label{eq:mass_energy}
 &&  \nabla_{\mu} T^{\mu}_{\nu} = -\mathcal{S}_{\nu} \\
     \label{eq:maxwell}
   && \nabla_{\nu} F^{*\mu \nu} = 0 
\eea
in a spherical-like Kerr-Schild coordinates ($t,\ r, \ \theta, \ \phi$) with $G=c=M_{BH}=1$.
Equations (\ref{eq:mass}), (\ref{eq:mass_energy}), and (\ref{eq:maxwell}) describe the conservation of particle number, conservation of energy-momentum, and source-free Maxwell equations, respectively.
%, as well as the no-magnetic monopole constraint. 
We maintain the divergence-free condition of the magnetic fields using a `Constrained Transport' (CT) \citep{Gardiner2005, White2016} update of the face-centred magnetic fields.

The stress-energy tensor is given by 
 \be
 \label{eq:stress_tensor}
     T^{\mu \nu } = \left(\rho h + b^2 \right) u^{\mu} u^{\nu} + \left( p_{\rm gas} + \frac{b^2}{2} \right) g^{\mu \nu} - b^{\mu} b^{\nu}
 \ee
 and the dual of the electromagnetic field tensor is
 \be
 \label{eq:dual}
  F^{* \mu \nu} = b^{\mu} u^{\nu} - b^{\nu} u^{\mu}
  \ee
where $\rho$ as the comoving rest mass density, $p_{\rm gas}$ is the comoving gas pressure, $u^{\mu}$ is the coordinate frame 4-velocity,  $\Gamma=5/3$ is the adiabatic index of the gas, $h=1 + \Gamma/(\Gamma -1) p_{\rm gas}/\rho$ is comoving enthalpy per unit mass and $g_{\mu \nu}$ is the metric tensor in Kerr-Schild coordinates. The term $\mathcal{S}_{\nu}$ represents an optically thin cooling term which will be discussed in detail in section \ref{sect:cooling_fun}.  The 3-magnetic field in the coordinate frame, $B^{i}= F^{*i0}$ is related to the 4-magnetic field $b^{\mu}$ as
  \bea
  \label {eq:b0}
   &&   b^t = g_{i \mu} B^{i} u^{\mu},\\
   \label{eq:b1}
   &&  b^i  = \frac{B^i + b^t u^i}{u^t}.
  \eea
 For diagnostics, we also use magnetic field components ($B_{\bar{r}}, \ B_{\bar{\theta}}, \ B_{\bar{\phi}}$) defined in a spherical-polar like quasi-orthonormal frame as
 \be
 \label{eq:B_r_th_phi}
 B_{\bar{r}} = \sqrt{g_{rr}}B^r, \ B_{\bar{\theta}} = \sqrt{g_{\theta \theta}}B^{\theta}, \ B_{\bar{\phi}} = \sqrt{g_{\phi \phi}} B^{\phi}.
 \ee

 We use the GRMHD code {\tt Athena++} \citep{White2016, Stone2020} to perform the simulations.
We employ the HLLE solver \citep{Einfeldt1988}  with a third-order piecewise parabolic method (PPM; \citealt{Colella1984}) for spatial reconstruction. For time integration, a second-order accurate van Leer integrator  is used with the CFL number 0.3. 
All the length scales and time scales  in this work are expressed in units of the gravitational radius $r_g=GM_{BH}/c^2$ and $t_g=r_g/c$ respectively unless stated otherwise.

\subsection{Cooling function}
\label{sect:cooling_fun}
We aim to simulate a radiatively efficient geometrically thin disk around a black hole. However, we initialize the simulation with a geometrically thick Fishbone-Moncrief torus (\citealt{Fishbone1976}). Additionally, viscous heating contributes to the disk's puffing up. Hence, to make the disk thin and radiatively efficient, we use an {\em ad hoc}  cooling function (\citealt{Noble2009,Fragile2012a}), which adds a radiative loss term $\mathcal{S}_{\mu} = \mathcal{S}u_{\mu}$ to the energy-momentum equation (\ref{eq:mass_energy}) after a cooling switch-on time $t_{\rm switch}=10^4$. The cooling function  
\begin{equation}
  \setlength{\arraycolsep}{0pt}
  \mathcal{S} = t_{\rm cool}^{-1} \  \ U_g \   \left\{ \begin{array}{ l l }
    0, \quad & Y<1; \\
    Y-1, \quad &   1 < Y < Y_{\rm crit}; \\
    Y_{\rm crit} -1, \ \ \   \quad & Y>Y_{\rm crit}.
  \end{array} \right.
\end{equation}
is the rate at which energy is radiated per unit volume in the fluid frame. Here, the dimensionless number $Y=p_{\rm gas}/\rho T_{\ast}$ is the ratio of the gas temperature to the target temperature $T_{\ast}$ which is defined by the target aspect ratio $\epsilon^{\ast}_{\rm trgt}=H^{\ast}_{\rm th}/R$ and given by
\be
\label{eq:target_temp}
T_{\ast}  = \left[\epsilon^{\ast}_{\rm trgt} R \Omega_K \right]^2,
\ee
 with $U_g=p_{\rm gas}/(\Gamma-1)$ as internal energy of the gas and $H_{\rm th}=c_s/\Omega_K$ is the thermal scale-height. We choose $\epsilon^{\ast}_{\rm trgt}=0.1$ and $Y_{\rm crit}=4$.

The cooling time $t_{\rm cool}$ is given by
\be
\label{eq:cooling_time}
   t_{\rm cool}^{-1} = q_{\rm cool} \ \Omega_K \ \mathcal{F} (\theta),
\ee
where $\Omega_K$ is the relativistic orbital frequency and $\mathcal{F}(\theta)=\sin^{8} \theta$ is a modulation function used to make cooling inefficient at high latitudes of the computational domain. Hence, gas essentially evolves adiabatically in the jet and the region at high latitudes far from the disk. The parameter $q_{\rm cool}$ defines the efficiency of cooling: the larger its value, the shorter is the cooling time and vice-versa. We choose $q_{\rm cool}=10.0$.

\subsection{Initial Conditions} 
 \label{sect:method_init}

\begin{figure}
\centering
 \includegraphics[width=.47\textwidth]{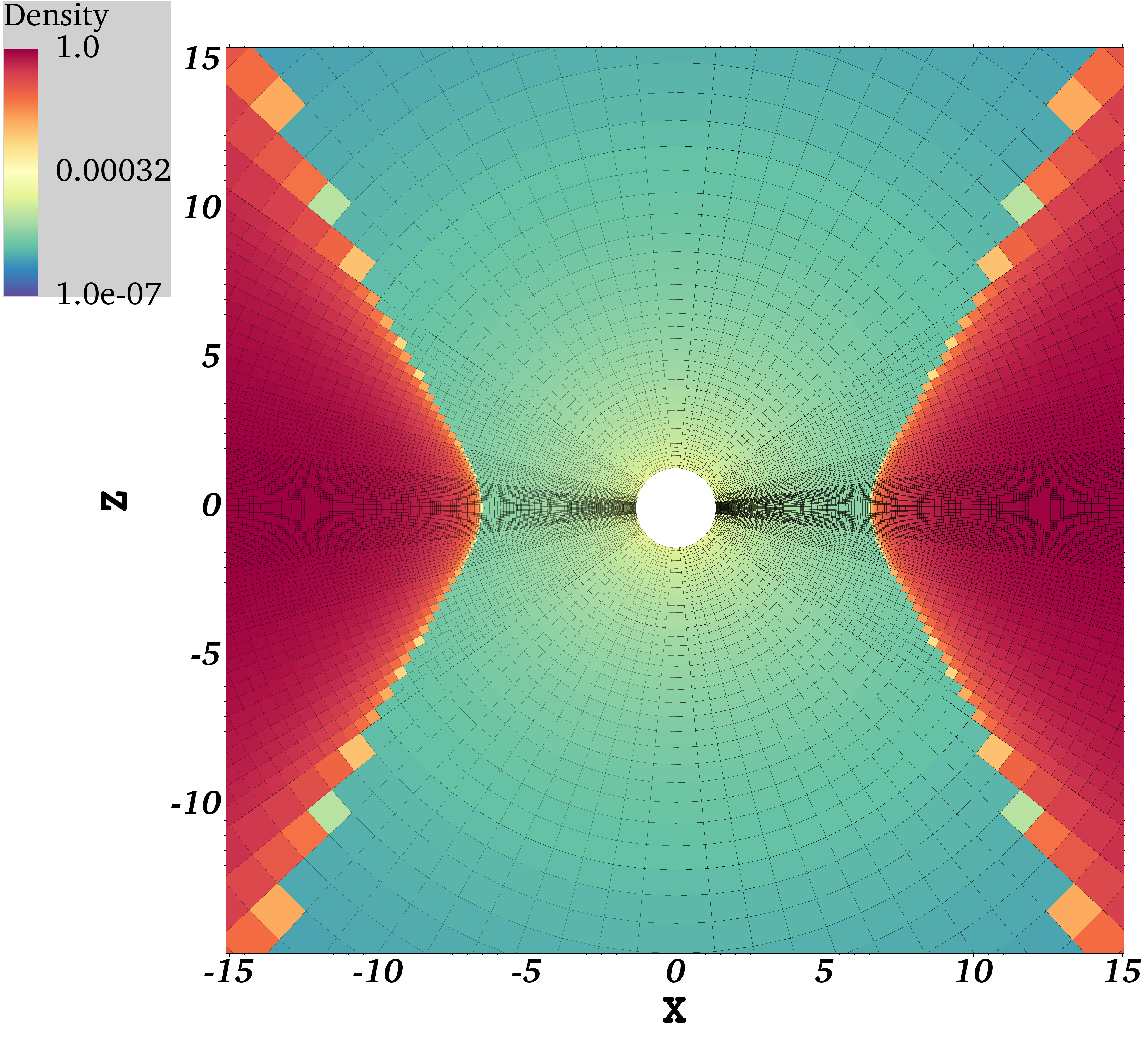}
\caption{Initial condition and grid set up in the poloidal plane. Colour shows the density, grid structures are depicted by the transparent lines.}
\label{fig:init_sim_a9}
\end{figure}
 
We start with a Fishbone-Moncrief (FM) torus (\citealt{Fishbone1976}) as shown in Fig.~\ref{fig:init_sim_a9} for the fiducial run with $a=0.9$. We choose different sets of torus parameters (see  Table \ref{tab:sim_tab}) for different BH spin $a$ such that the torus size approximately remains similar. We note that for the retrograde run $a9m$, we essentially anti-align the disk's angular velocity direction with respect to the BH spin. Nevertheless, we refer to this run as $a=-0.9$.

We initialize the magnetic field with a vector potential  which is non-zero inside the region $r \in \left [r_{\rm B,min}, r_{\rm B,max} \right]$,  $\theta \in \left[\theta_{\rm B,min}, \theta_{\rm B,max} \right]$, and $\phi \in \left[0, 2\pi \right]$ and defined by
\be
\begin{split}
A_{\phi} = C_{B} \ \ {\rm max} \left( p_{\rm gas} - p_{\rm gas, c}, \ 0.0 \right)^{ 1/2} \  \ r  \sin \theta \\
            \sin \left ( \pi  \mathcal{R} \right)
             \sin \left( \pi   \Theta \right)   
\end{split}
\ee
where
$\mathcal{R}=(r-r_{\rm B,min})/(r_{\rm B,max} - r_{\rm B, min})$ and $\Theta =(\theta-\theta_{\rm B,min})/(\theta_{\rm B,max} - \theta_{\rm B, min})$ . Here, we choose $p_{\rm gas,c}=10^{-8}$, $r_{\rm B,max}=2000$, $\theta_{\rm B,min}=\pi/6$ and $\theta_{\rm B,max}=\pi-\theta_{\rm B,min}$. The values of $r_{\rm B,min}$ are tabulated in Table \ref{tab:sim_tab}. We choose the normalization constant $C_B=75$ for all the runs. The choice of $C_B$ gives rise to similar initial magnetic field strength inside the torus with initial plasma $\beta_{\rm ini}=\int_{\rm torus} p_{\rm gas} \rho dV/\int_{\rm torus} p_{\rm mag} \rho dV \approx 70$, where $dV=\sqrt{-g} dr d\theta d\phi$ is volume element.

\begin{deluxetable}{cccccc}
\tablenum{1}
\tablecaption{Details of the simulations with BH spin $a$, inner boundary $r_{\rm in}$ of the computational domain, inner edge of the torus $r_{\rm edge}$, location of pressure maximum of the torus $r_{\rm max}$, magnetic field cut-off radius $r_{\rm B,min}$. \label{tab:sim_tab}}

 \tablehead{
  \colhead{Name} & \colhead{$a$} & \colhead{$r_{\rm in}$}  & \colhead{$r_{\rm edge}$}   & \colhead{$r_{\rm max}$}	 & \colhead{$r_{\rm B,min}$}  
 }
\decimalcolnumbers
\startdata
 a0     & 0    & 1.87   & 6.5   & 17.99     & 8.0    \\
 a3     & 0.3  & 1.83   & 6.5   & 16.85     & 8.0    \\
 a5     & 0.5  & 1.75   & 6.5   & 16.30     & 8.0    \\
 a9     & 0.9  & 1.34   & 6.5   & 15.30     & 8.0    \\
 a99    & 0.99 & 1.06   & 6.5   & 15.14     & 8.0   \\
 a9m    & -0.9 & 1.34   & 9.0   & 23.58     & 9.2   \\
 \enddata

\end{deluxetable}

\subsection{Numerical setup}
\label{sect:method_setup}
Our computational domain spans $r \in [r_{\rm in}, 1000], \ \theta \in [0, \pi], \ \phi \in [0, 2\pi]$, where we keep the inner boundary $r_{\rm in}$ (see Table \ref{tab:sim_tab}) inside the outer event horizon $r_H=m+\sqrt{m^2 -a^2}$ with eight grid points between inner boundary and the event horizon. This allows a causally disconnected inner boundary. Radial grids are spaced logarithmically, while meridional grids are compressed towards the mid-plane (for details see \citealt{Dhang2023}) such that  the meridional cell widths
in the polar region is approximately twice that at  the mid-plane ($\theta=\pi/2$). Uniform grids are employed in the azimuthal direction. We use a root grid resolution $96 \times 48 \times 64$, giving rise to $\Delta r : r \Delta \theta : r \Delta \phi = 1.5: 1: 2 $ at the equator in the Newtonian limit. We use three levels of static refinement (for a visual see Fig.~\ref{fig:init_sim_a9}) to improve the resolution in each direction by a factor of eight. The first, second and third levels of refinement are employed in the regions $\mathcal{L}_1: r \in [r_{\rm in}, r_{L1}], \theta  \in [55^{\circ},125^{\circ}], \phi \in [0, 2\pi]$; $\mathcal{L}_2: r \in [r_{\rm in}, r_{L2}], \theta  \in [75^{\circ},105^{\circ}], \phi \in [0, 2\pi]$;  $\mathcal{L}_3: r \in [r_{\rm in}, r_{L2}], \theta  \in [83^{\circ},97^{\circ}], \phi \in [0, 2\pi]$; respectively. The choice of $r_{L1}$, $r_{L2}$ and $r_{L3}$ depends on the BH spin with $80<r_{L1}<115$, $50<r_{L2}<70$, $30<r_{L3}<40$. Thus at the disk-midplane, we have an effective resolution of $768 \times 768 \times 512$ considering the fact that  meridional grids are compressed towards the mid-plane by a factor of two. We found that this choice of resolution is adequate for convergence. For a detailed discussion on convergence see Appendix \ref{sect:convergence_appendix}.

We use a pure inflow boundary condition ($u^{r} \leq 0$) at the radial inner boundary, while at the radial outer boundary, primitive variables are set according to their initial radial gradients. Magnetic fields in the ghost zones are copied from the nearest computation zone both at inner and outer radial boundaries.  Polar and periodic boundary conditions are used at the meridional and azimuthal boundaries, respectively.

We use the following floors on density and pressure to stabilize the code: $\rho_{\rm floor} = {\rm max} \left[ 10^{-4}  r^{-3/2}, 10^{-8} \right]$ and  $p_{\rm gas,floor} = {\rm max} \left[ 10^{-6}  r^{-5/2}, 10^{-10} \right]$. We also constrain the following variables: $\beta > 10^{-3}$, magnetization $\sigma = 2 p_{\rm mag}/\rho < 100$ and Lorentz factor $\gamma < 50$.

\subsection{Diagnostics}
\label{sect:diagnostics}
We now list and define the diagnostics used in the rest of the paper. The mass accretion rate is
\be
\label{eq:mdot}
\dot{M}(r,t) = - \int \rho u^{r} dS_r,
\ee 
where $dS_r = \sqrt{-g} \ d\theta \ d\phi$ is the area element of the $r=constant$ surface.

The energy efficiency associated with the gas (matter + electromagnetic), $\eta_g$, is defined as
\be 
\label{eq:eta_gas}
\begin{split}
\eta^{g}(r,t) &= - \frac{1}{\dot{M}_H (r,t)} \int_{S} \left(\rho u^r + T^{r}_{t}\right) \ dS_r \\
        &= \eta^{M} + \eta^{EM},
\end{split}
\ee 
where $T^{r}_{t} = \left(T^{r}_{t} \right)^{M} + \left(T^{r}_{t} \right)^{EM} $ with $\left(T^{r}_{t} \right)^{M} =\rho h u^{r}u_{t}$ and $\left(T^{r}_{t} \right)^{EM} = 2p_{\rm mag}u^{r}u_{t} - b^{r}b_{t}$. The quantities $\eta^{M}$ and $\eta^{EM}$ are the efficiencies associated with matter and electromagnetic fields, respectively, and $\dot{M}_H (t)$ is the average (from $t-\Delta t/2$ to $t+\Delta t/2$ with $\Delta t= 400$) mass accretion rate at the BH horizon. The quantity $\eta^{EM}$ at any radius $r$ is defined as
\be
\label{eq:eta_em}
\eta^{EM}(r,t) =  \frac{P^{EM}_{\rm out}}{\dot{M}_H}=- \frac{1}{\dot{M}_H} \int_{S} \left(T^{r}_{t}\right)^{EM} \ dS_r.
\ee 
Here, it must be noted that the surface integrals in equations \ref{eq:eta_gas} and \ref{eq:eta_em} 
are generally carried out over the entire sphere. However, we sometimes limit it to a specific range of $\theta$, depending on the purpose. For example, to calculate the jet efficiency, the integration is performed at $r=50$ with $\theta$-integration covering only the highly magnetized and relativistic polar regions following \cite{EHT_V_2019}. Therefore, jet efficiency is calculated as
\be
\label{eq:eta_j}
\eta^{EM}_{j} =- \frac{1}{\dot{M}_H} \int_{r=50} \digamma \left(T^{r}_{t}\right)^{EM} \ dS_r,
\ee 
where $\digamma$ is a step function whose value is unity only in regions where $\sigma >1$ and 
$\gamma v > 1$; otherwise its value is zero.

The magnetization of the accretion flow in a global simulation is often characterized by the amount of magnetic flux (normalized by the mass accretion rate) threading each hemisphere of the BH horizon or, more specifically, by the `MAD parameter' defined by
\be 
\label{eq:mad_param}
\phi_{BH}(t) = \frac{\sqrt{4 \pi}}{2 \sqrt{\dot{M}_{H}}} \int |B^{r}(r_H,\theta,\phi)| \  dS_r.
\ee 

To characterize the disk thickness, in addition to the thermal scale height $H_{\rm th}=c_s/\Omega_K$ which describes the gas temperature in the disk mid-plane, we additionally define two other scale heights describing the vertical density distribution of gas. The density-weighted scale heights are defined as
 \be
 \label{eq:h_r}
 H/r =  \left( \frac{\int \la \bar{\rho}(r,\theta) \ra  \  |\frac{\pi}{2} -\theta|^{n} \ \sqrt{-g} \ d\theta}  {\int \la  \bar{\rho}(r,\theta) \ra  \ \sqrt{-g} \  d\theta} \right)^{\frac{1}{n}},
\ee
with $n=1$ and $n=2$ describing disk aspect ratios $\epsilon=H/r$ and $\epsilon_{ \rm rms}=H_{\rm rms}/r$ respectively.  Here, $\bar{...}$ and  $\la ... \ra $ represent the azimuthal and time average, respectively. For a disk with a Gaussian vertical density profile, $\epsilon=\sqrt{2/\pi} \epsilon_{\rm th}$.

\section{Fiducial run with $\texorpdfstring{\MakeLowercase{a}}{a}=0.9$ }
\label{sect:fiducial_run}

Here, we describe the basic features of our fiducial simulation of a thin disk ($\epsilon_{\rm th}=0.1$) around a BH with spin $a=0.9$.

\subsection{Flow evolution: transition from hot-thick to cold-thin MAD}
\label{sect:hot_to_cold_mad}
\begin{figure*}
\centering
 \includegraphics[scale=0.55]{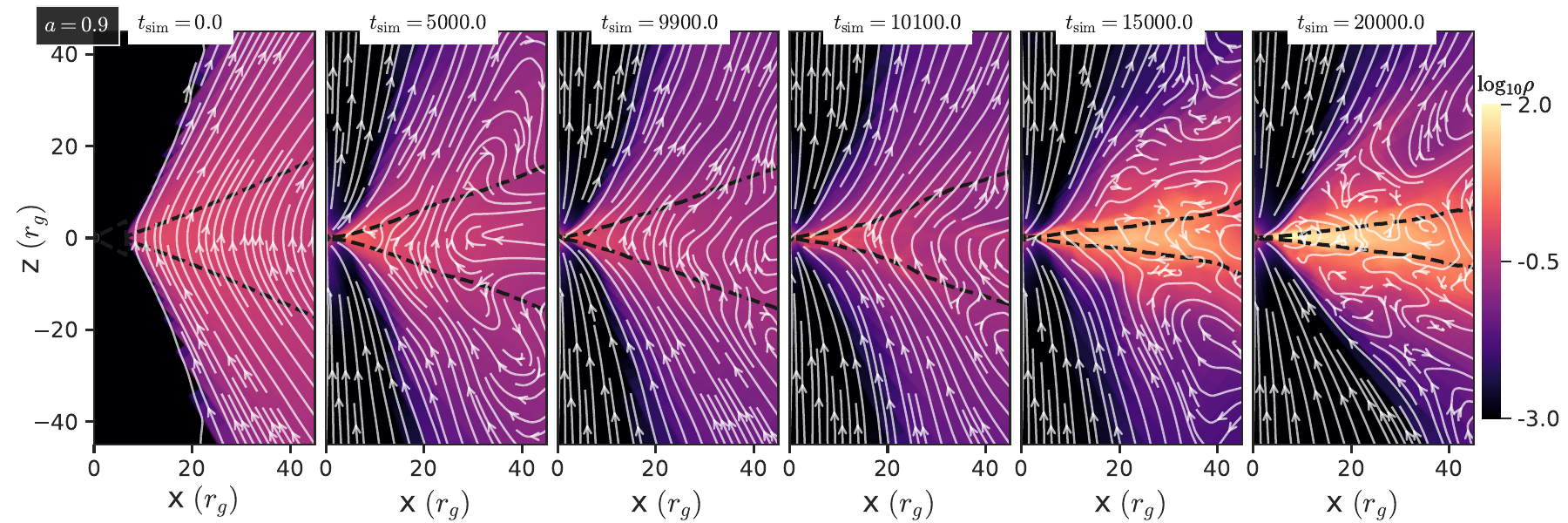}
\caption{Evolution of the initial FM torus in the poloidal plane for the fiducial run with $a=0.9$. Color and streamlines show azimuthally averaged density $\bar{\rho}(r,\theta)$ and poloidal magnetic field $\bar{\mathbf{B}}_{p} (r,\theta) = \bar{B}_r (r,\theta) \hat{r} + \bar{B}_{\theta} (r,\theta) \hat{\theta}$, respectively. The first three panels describe the evolution of the non-radiative torus, the end state of which is a quasi-stationary geometrically thick hot flow. The last three panels depict the torus evolution after cooling is switched on at time $t=t_{\rm switch}=10^4$. Due to cooling, the torus loses its pressure support and forms a thin disk of aspect ratio $\epsilon = H/r$, depicted by black dashed lines in each panel.   }
\label{fig:density_a9}
\end{figure*}

\begin{figure*}
\centering

\includegraphics[scale=0.55]{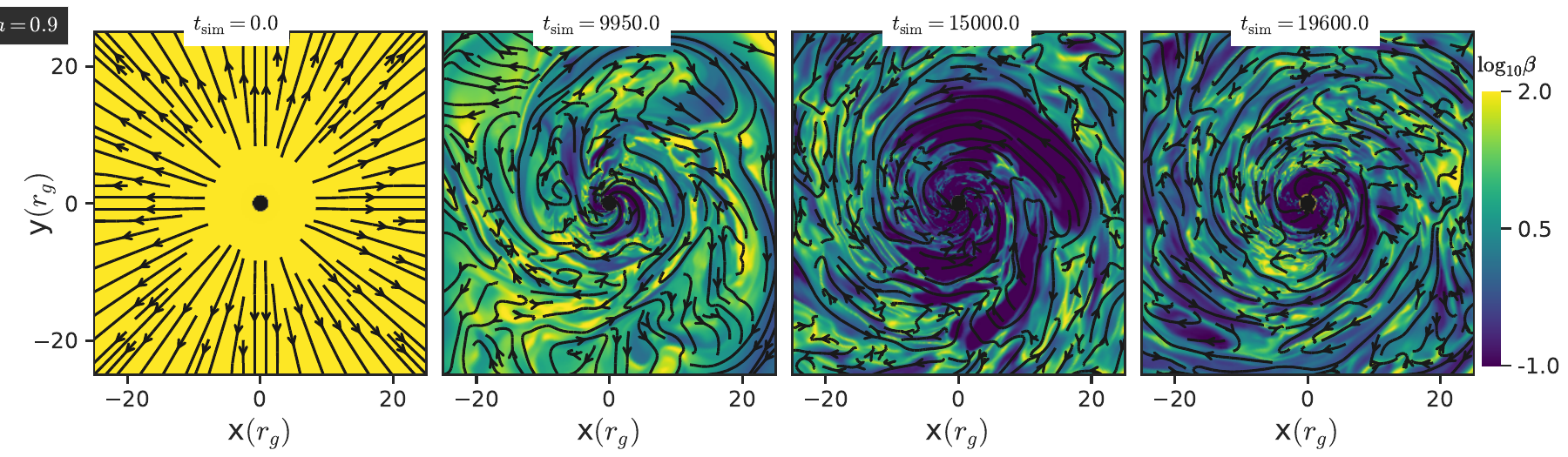}
\caption{Snapshots of the plasma beta,  $\beta=p_{\rm gas}/p_{\rm mag}$,  at the disk mid-plane for the fiducial run with $a=0.9$. Color represents $\beta$; streamlines describe magnetic fields.  The simulation starts with a purely poloidal magnetic field, but the toroidal component grows due to strong shear and becomes dominant during the later evolution, especially as the accretion flow transitions (starting at $t=10^4$) from a geometrically thick hot flow to a geometrically thin Keplerian disk.  }
\label{fig:beta_a9}
\end{figure*}

\begin{figure}
\centering

\includegraphics[scale=0.55]{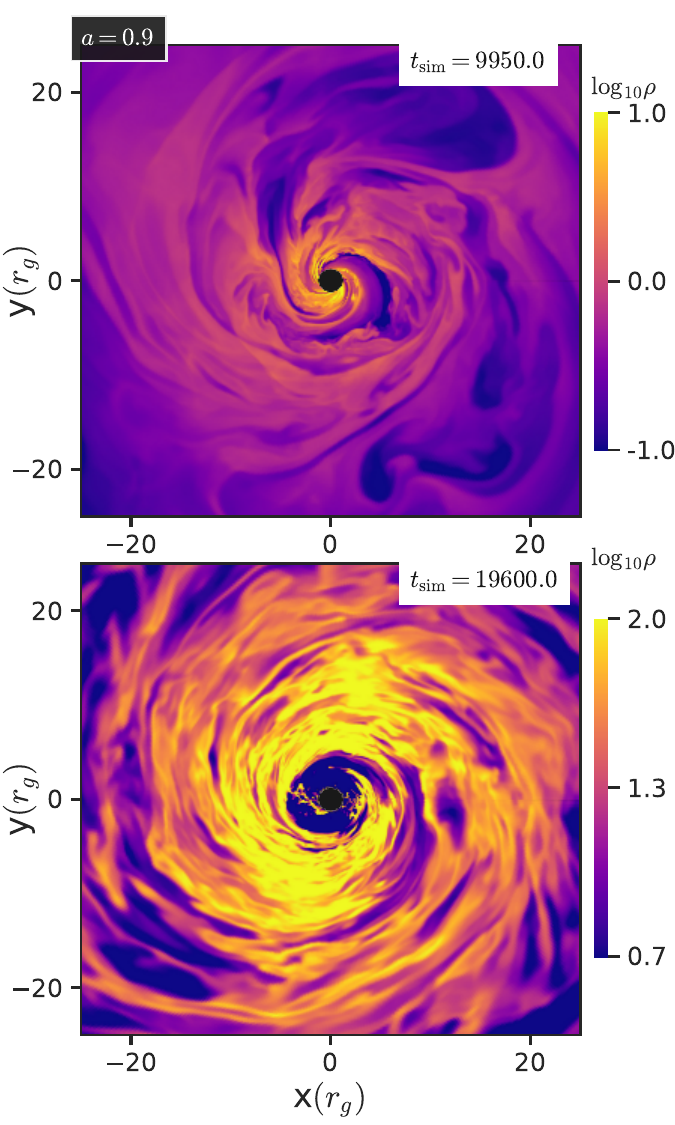}
\caption{Snapshots of the rest mass density $\rho$ at the disk mid-plane for geometrically thick hot (at $t=9950$) and geometrically thin cold (at $t=19600$) MADs, respectively, for the fiducial run with $a=0.9$.
Both hot and cold MADs feature the presence of low-density cavities which are highly magnetized (see Fig. \ref{fig:beta_a9}) and related to magnetic flux eruptions.}
\label{fig:rho_xy}
\end{figure}

Figure \ref{fig:density_a9} illustrates the transformation from a geometrically thick, hot accretion flow to a geometrically thin, cold accretion flow, with snapshots of density and magnetic fields at various times. The color scale represents density $\bar{\rho}(r,\theta)$ and poloidal magnetic fields $\bar{\mathbf{B}}_{\bar{p}} (r,\theta) = \bar{B}_{\bar{r}} (r,\theta) \hat{r} \ + \  \bar{B}_{\bar{\theta}} (r,\theta) \hat{\theta}$ are depicted by streamlines. The dashed black lines indicate the disk thickness, defined by the disk-scale height $\epsilon=H/r$.

Initially, a geometrically thick torus threaded by a poloidal magnetic field, is set up, as shown in the first panel of Figure \ref{fig:density_a9}.  We evolve it adiabatically and after a sufficient time, the region near the black hole becomes saturated with magnetic flux, leading to a MAD state (\citealt{Tchekhovskoy2011}), characterized by strong jets and a geometrically thick, hot configuration (second and third panels of Figure \ref{fig:density_a9}).

To transition to a geometrically thin, cold MAD, the cooling term, described in Section \ref{sect:cooling_fun}, is activated at $t=t_{\rm switch}=10^4$. The fourth panel of Figure \ref{fig:density_a9} shows the accretion flow shortly after cooling is initiated. As the disk loses vertical pressure support, it becomes thinner, with the inner region cooling first due to the prescribed cooling time $t_{\rm cool} \propto r^{3/2}$ (see equation \ref{eq:cooling_time}). Over time, the outer disk region also cools, and a larger fraction of the gas settles down around the disk-midplane and forms a geometrically thin, cold MAD attaining the targeted thermal disk-thickness, $\epsilon_{\rm th}=0.1$, as shown explicitly in the latter part of this section.

Figure \ref{fig:beta_a9} describes the azimuthal distribution of plasma $\beta=p_{\rm gas}/p_{\rm mag}$ and magnetic fields ($B_{\bar{r}}$, $B_{\bar{\phi}}$) in the disk-midplane. The first and second panels in Figure \ref{fig:beta_a9} describe the initial condition and the quasi-stationary phase of the hot MAD. The last two panels of Figure \ref{fig:beta_a9} illustrate the azimuthal structure of plasma-$\beta$ and magnetic fields for the cold thin disk. Starting from a weak poloidal field with $\beta \gg 1$, the magnetic field gets amplified due to shear and becomes dominated by the toroidal component. Flux freezing and possibly MRI (\citealt{Begelman2022}) are also two important factors in determining the amplification of the magnetic field. While the hot MAD saturates at $\beta_{av} \sim 50$ in the disk mid-plane, the cold thin disk has a lower $\beta_{av} \sim 1$ in the quasi-stationary phase. We also noticed the presence of low-density cavities, which are highly magnetized, as evident from the comparison of plasma $\beta$ and rest mass density $\rho$ in Figs.~\ref{fig:beta_a9} and \ref{fig:rho_xy}, both before and after switching on the cooling. These cavities are characteristic features of a MAD (\citealt{Porth2021, Ripperda2022, Scepi2022}) and are associated with the magnetic flux eruptions (\citealt{Begelman2022}).
This is further illustrated in the $\phi_{\rm BH}-t$ plot presented in section \ref{sect:madness}. 
Finally, by the end of the simulation at $t=2 \times 10^4$ with $a=0.9$ we observed that the thin MAD had reached a quasi-stationary state within $r\approx 7$ which we denote as the inflow equilibrium radius $r_{\rm eq}$. This characteristic radius, marking the extent of the quasi-stationary accretion flow, is typically defined as the radial distance over which the time-averaged mass accretion rate $\dot{m}(r)$ remains constant.

\subsection{Characterizing the disk thickness}

\begin{figure*}
%\centering
\gridline{  \fig{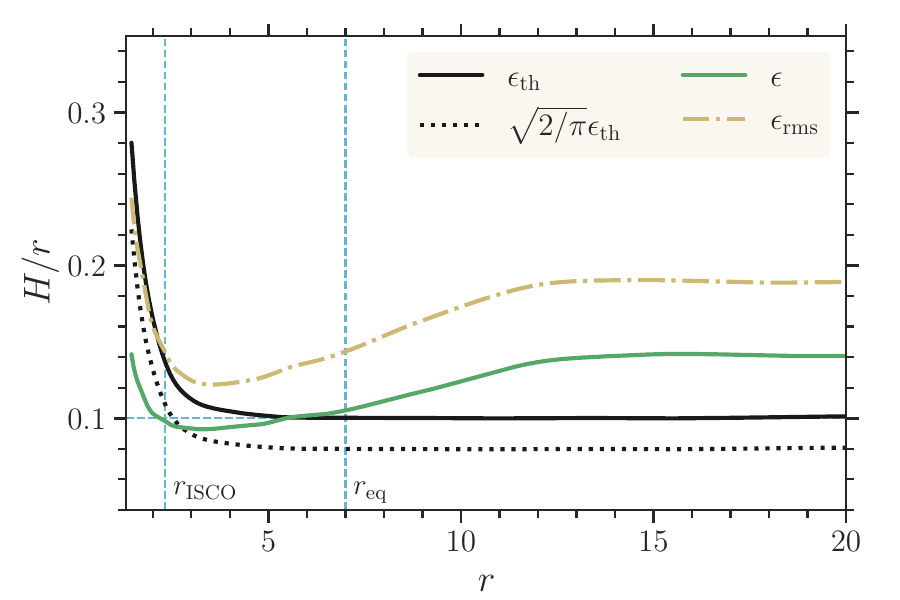}{0.49\textwidth}{(a)}
            \fig{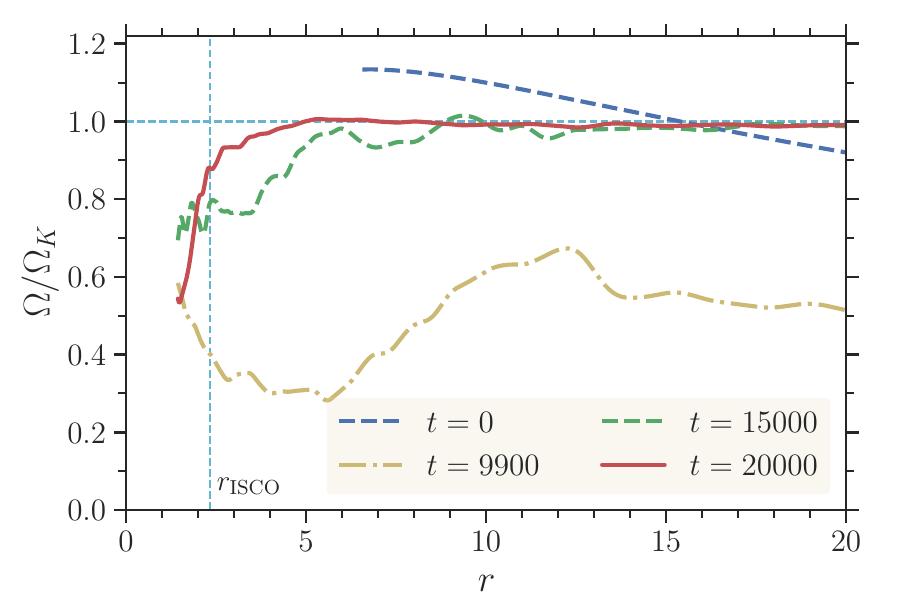}{0.49\textwidth}{(b)}        
          }
\caption{Left-hand panel: Radial variation of time-averaged thermal scale-height $\epsilon_{\rm th}=H_{\rm th}/r$, density-weighted scale-height $\epsilon=H/r$ and rms scale-height $\epsilon_{\rm rms}=H_{\rm rms}/r$ in the quasi-stationary state. The time average is done over $t=18000-20000$. The horizontal cyan dashed line denotes the targeted value of the thermal scale height in the cooling function, while the vertical cyan dashed lines depict the locations of the ISCO ($r_{\rm ISCO}$) and inflow equilibrium radius ($r_{\rm eq}$), respectively.  Right-hand panel: Radial distribution of the normalized angular velocity $\Omega/\Omega_K$ of the fluid in the disk midplane at different times for the run a9. The horizontal and vertical cyan dashed lines denote the Keplerian angular velocity and the location of the ISCO, respectively. }
\label{fig:h_r_omega_a9}
\end{figure*}

We have qualitatively described the formation of a cold, thin disk from a geometrically thick hot MAD in section \ref{sect:hot_to_cold_mad}. In this section, we quantify the disk thickness and Keplerianity of the angular velocity characterizing a thin disk.

The left-hand panel of Fig.~\ref{fig:h_r_omega_a9} shows the radial profiles of average (average over $t=18000-20000$) disk scale-height using different definitions: thermal disk height $\epsilon_{\rm th}=H_{\rm th}/r$, density-weighted scale height $\epsilon=H/r$ and rms scale-height $\epsilon_{\rm rms }=H_{\rm rms}/r$ (see equation \ref{eq:h_r}). We find that the cooling function used in our simulation is pretty efficient in attaining the target aspect-ratio of $\epsilon_{\rm th}= \epsilon^{\ast}_{\rm trgt} =0.1$ for the region outside the ISCO. However, as the ISCO is approached, disk thickness rapidly shoots up, probably due to the inefficient cooling in the plunging region (\citealt{Noble2009}). We also noticed that the density-weighted scale-height $\epsilon$ is larger than the thermal one, $\epsilon_{\rm th}$ at all radii, except for the region close to the plunging region;  and is also different from what is expected in a standard disk with Gaussian vertical density profile for which $\epsilon = \sqrt{2/\pi} \epsilon_{\rm th} = 0.798 \epsilon_{\rm th}$. This is similar to what has been observed in \cite{Scepi2024}, who found that for a thin MAD with $\epsilon_{\rm th} < 0.125$, turbulent magnetic pressure becomes the dominant factor in determining the vertical structure instead of thermal pressure, resulting in a thicker disk than the thermal scale-height. The rms scale-height $\epsilon_{\rm rms}$ is always larger than $\epsilon$, similar to earlier MAD simulations of similar scale-height (\citealt{Avara2016}).

The right-hand panel of Fig.~\ref{fig:h_r_omega_a9} shows the radial profiles normalized (by the Keplerian value) azimuthally averaged angular velocity at the disk-midplane at different times. We start with a constant angular momentum torus at $t=0$, which is Keplerian at the density maximum $r=r_{\rm max}$, sub-Keplerian for $r>r_{\rm max}$ and super-Keplerian for $r<r_{\rm max}$. As the flow evolves, the angular velocity forgets its initial distribution and becomes around 50\% of the Keplerian value, which is
a characteristic of a pressure-supported hot MAD (\citealt{Narayan2012, Begelman2022}). As we switch on the cooling at $t=10^4$, the accretion flow loses its thermal pressure support. It becomes quasi-Keplerian, and the centrifugal barrier supports the radial gravity in the thin disk.

\subsection{Characterizing the MAD-ness}
\label{sect:madness}
\begin{figure}
\centering
 \includegraphics[scale=0.55]{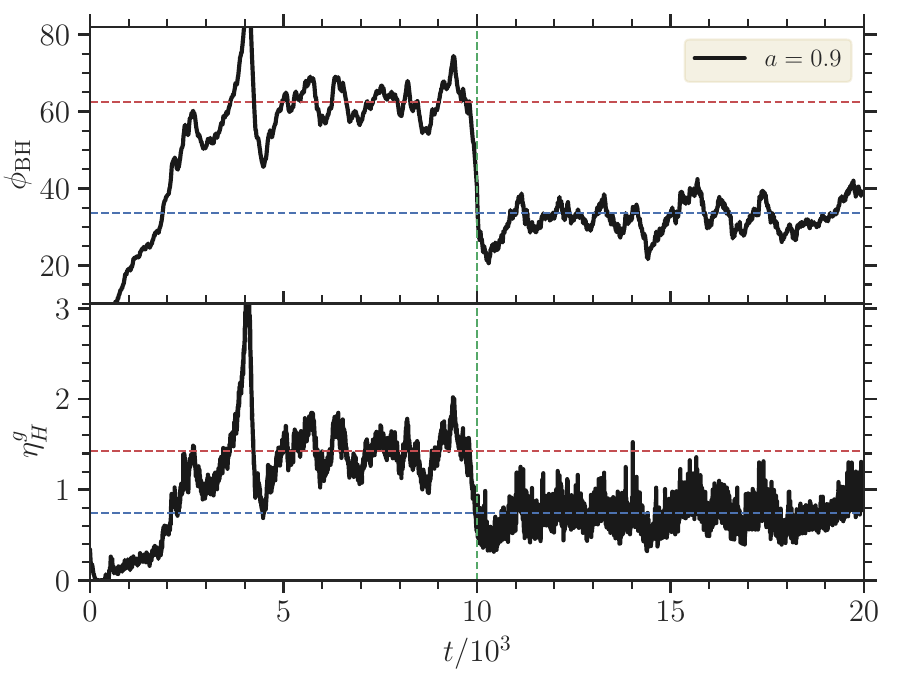}
\caption{Time series of the MAD parameter $\phi_{\rm BH}$ (top panel) and gas efficiency $\eta^{g}_H=\eta^{g}(r=r_H)$ (bottom panel) at the BH horizon for the run with $a=0.9$. The vertical green dashed line denotes when cooling is switched on. The horizontal red and blue dashed lines in each panel represent the time-averaged value of the respective quantities in the hot and cold phases of the flow. The time average is over $t=5500-10000$ for the hot flow and over $t=15500-20000$ for the cold disk. }
\label{fig:phi_bh_eta_a9}
\end{figure}

Ordered magnetic field structures, cavities, and sub-Keplerian angular velocity in the hot accretion flow in our simulation indicate that the accretion flow is magnetically arrested. However, earlier studies (e.g. \citealt{Tchekhovskoy2011,  McKinney2012, Narayan2022}) suggested that the most direct  method to characterize a MAD is to investigate the MAD parameter $\phi_{\rm BH}$ (equation \ref{eq:mad_param}). If its value is around or above the critical value $\approx 50$ in the hot accretion flow, it is described as in the MAD state (\citealt{McKinney2012}). In our simulation, as shown in the top panel of Fig. \ref{fig:phi_bh_eta_a9}, during the hot phase ($t<10^4$), the MAD parameter has an average value $\phi_{\rm BH} \approx 60$, implying a hot MAD. 

However, it has been found that the saturation level of $\phi_{\rm BH}$ is a strong function of disk thickness (see, e.g., \citealt{Scepi2024}). For our cold accretion flow, $\phi_{\rm BH}$ saturates around the value $\approx 32$, which is slightly larger than the value ($\phi_{\rm BH} \approx 25$) that has been found in earlier studies of similar disk thickness $\epsilon_{\rm th} = 0.1$, but with spin $a=0.5$. However, it should be noted that the MAD parameter also depends on the spin of the black hole, as we will see in the latter part of this paper in section \ref{sect:phi_bh_spin}  (also see \citealt{Narayan2022} describing spin dependence in hot accretion flows).  

Another feature of a MAD is that it is pretty efficient in liberating energy from  a spinning BH (\citealt{Tchekhovskoy2011, McKinney2012}). The bottom panel of Fig. \ref{fig:phi_bh_eta_a9} shows the time series of the gas efficiency  (equation \ref{eq:eta_gas}) calculated at the 
BH horizon, $\eta^{g}_H=\eta^{g}(r=r_H)$. The term $\eta^{g}_H$ can also be approximated as a combination of two terms,
\be
\label{eq:eta_g_h}
\eta^{g}_{H} \approx -\la Be (r=r_H) \ra \  + \ \int_{r=r_H} b^{r} b_{t} dS_r. 
\ee 
The first and second terms on the right-hand side of equation (\ref{eq:eta_g_h}) represent the average (over $\theta$ and $\phi$) Bernoulli number (\citealt{Penna2013_init, White2020})
\be
\label{eq:Be}
Be = -\frac{\rho u_t \ + \  \Gamma/(\Gamma-1) p_{\rm gas} u_t \ + \ 2p_{\rm mag} u_t }{\rho} \ - \  1
\ee 
and Poynting flux, calculated at the horizon, respectively. It is interesting to note that if the BH is accreting bound material ($Be<0$) it adds to the Poynting flux, increasing the efficiency, and vice-versa.

In the hot MAD, $\eta^{g}_H >1$ with an average value of $\approx 1.4$, while the average $\eta^{g}_H$ decreases to $\approx 0.75$  in the thin cold MAD. It is important to note that Poynting flux is the main contributor to $\eta^{g}_H$ in the hot flow. At the same time, the contribution from the Bernoulli term increases significantly in the cold disk, because of the accretion of the enormous amount of bound material by the BH. The value of $\eta^{g}_H>1$ dominated by Poynting flux in the hot flow indicates the presence of a Blandford-Zanjek (BZ) process extracting rotational energy of the black hole and powering the jets (\citealt{Blandford1977, Tchekhovskoy2011}). We find that the BZ process is also active in the cold MAD and is responsible for jet launching, as discussed in detail in the rest of the paper.  

In summary, our simulation utilizing an {\em ad-hoc} cooling function has successfully produced a cold, thin disk with a thermal scale-height of  $\epsilon_{\rm th}=0.1$, exhibiting quasi-Keplerian angular velocity. Furthermore, the flow is characterized by the presence of highly magnetized, low-density cavities, with a quasi-stationary $\phi_{BH} \approx 30$, indicating the formation of a thin cold MAD.

\section{Results for thin MADs} 
\label{sect:results_thin_mad}
In the previous section, discussing the fiducial run for $a=0.9$, we have shown that a geometrically thin cold MAD can be created by cooling a hot MAD. In this section, we discuss how energy extraction depends on BH spin energy.

\subsection{Blandford-Znajek process in the thin disk}
\label{sect:Bz_intro}
\begin{figure*}
\centering
 \includegraphics[scale=0.6]{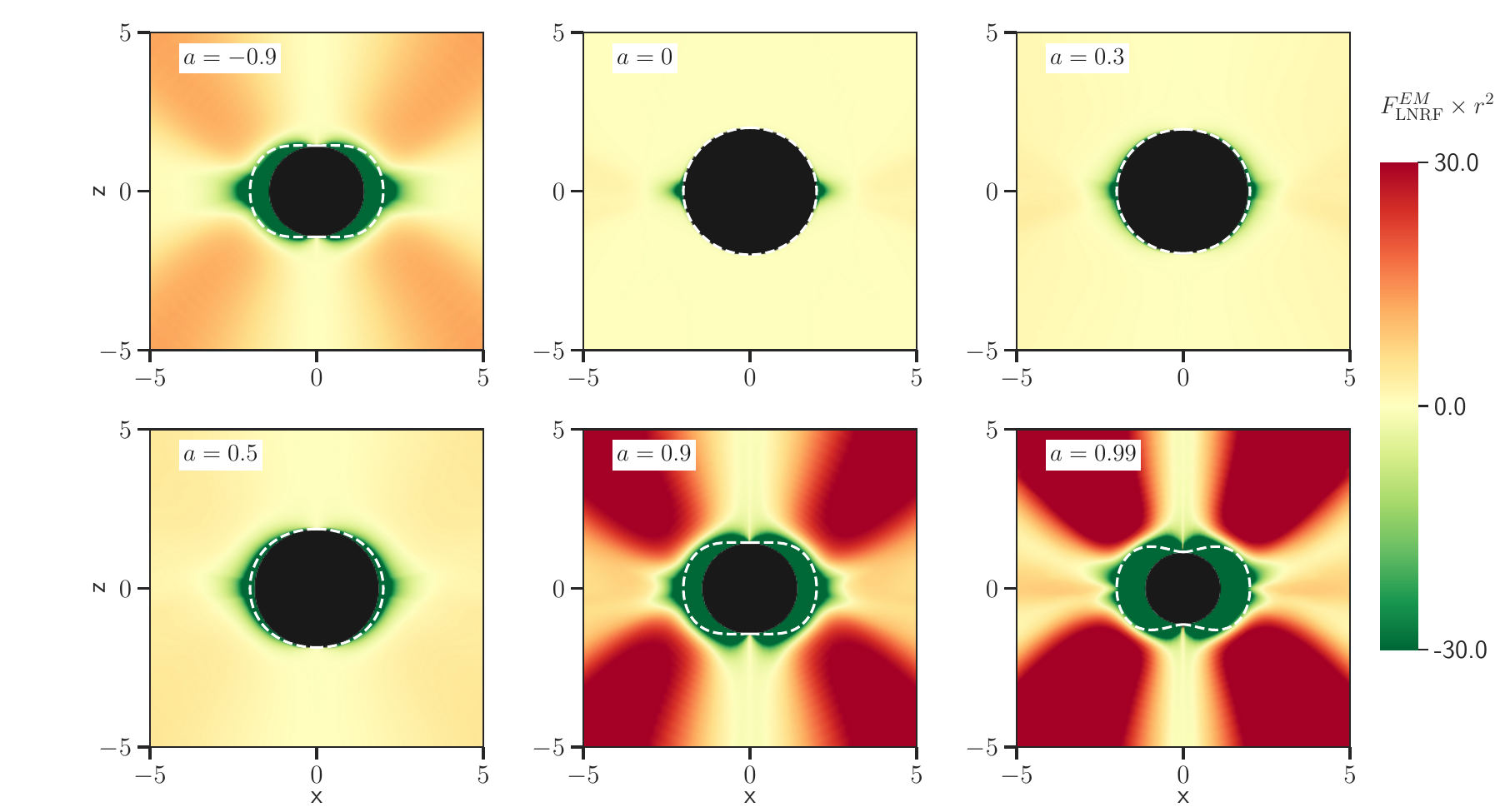}
\caption{Time-averaged and $\phi-$ averaged Poynting flux $F^{EM}_{\rm LNRF}=-T^{\hat{r}}_{\hat{t}}$ observed by a local static observer (in a locally non-rotating reference frame) for cold thin MADs
surrounding black holes with varying spins. The time average is taken over the interval $t=18000-20000$. The white dashed contour depicts the location of the ergosphere. Negative energy states within the inner light cylinder (ergosphere) are clearly visible, particularly for high spin cases, where there is a prominent separation between the event horizon and ergosphere. This qualitatively demonstrates the presence of the Blandford-Zanjek process around the spinning black holes, which is proposed to be the electromagnetic analogue of the Penrose process.}
\label{fig:poynting_zamo}
\end{figure*}

 Multiple previous studies have demonstrated that the BZ mechanism provides a good explanation for understanding the
extraction of BH spin energy (e.g. \citealt{Tchekhovskoy2011, McKinney2012, Penna2013}) with an efficiency (\citealt{Tchekhovskoy2010})
\be
\label{eq:BZ_eff}
\eta_{BZ} = \frac{\kappa}{4 \pi} \  \phi^2_{BH} \ \Omega^2_H,
\ee 
where
\be 
\label{eq:kappa}
\kappa = k_{\rm geo} \left(\frac{\Omega_F}{\Omega_H} \right) \left(1- \frac{\Omega_F}{\Omega_H} \right).
\ee
Here, $\Omega_F$ and $\Omega_H=a/2r_H$ are the angular frequency of the magnetic field lines and the BH, respectively. The parameter $k_{\rm geo}$ depends on magnetic fields geometry, particularly, $k_{\rm geo} \approx 2/3\pi$ and $k_{\rm geo} \approx 1.66/3\pi$ for 
split-monopole and paraboloidal field, respectively (\citealt{Blandford1977}).

The BZ process can provide more than 100\% efficiency, exceeding accreting rest mass energy provided that a hot MAD surrounds a highly spinning BH. We find an efficiency of 140\% for the hot MAD and black hole spin $a=0.9$. However, the efficiency goes down to 75\% in the case of a thin MAD for the same spin mainly because of the reduction in MAD parameter $\phi_{\rm BH}$, the most important factor in the BZ process. However, before quantifying the similarities/differences
between the energy extracted and the BZ prediction, we qualitatively demonstrate the BZ mechanism's presence in our spinning black hole simulations.

The BZ process is considered to be the electromagnetic analogue of the Penrose process \citep{Penrose1969} extracting BH spin energy.  In the BZ scenario, an outward Poynting flux in the coordinate frame (e.g., in Boyer-Lindquist or Kerr-Schild) appears inward propagating inside the inner light cylinder (here, the ergosphere of the BH) to a stationary observer in a locally non-rotating frame (LNRF; \citealt{Bardeen1972}). This occurs because the LNRF observer will see the electric fields reverse direction at the ergosurface (\citealt{Blandford1977, Znajek1977}). Fig.~\ref{fig:poynting_zamo} shows the distribution of azimuthally and time-averged Poynting flux $F^{EM}_{\rm LNRF}=-\left(T^{\hat{r}}_{\hat{t}}\right)^{EM}$ calculated in a LNRF in the poloidal plane. Here, 
\be 
u^{\hat{\mu}}=e^{\hat{\mu}}_{\nu} \  u^{\nu}_{BL}
\ee 
is the four-velocity in the LNRF; $e^{\hat{\mu}}_{\nu}$ is the orthonormal
tetrad ( for details see \citealt{Bardeen1972}) carried by the LNRF observer, and $u^{\mu}_{BL}$ is the four-velocity in Boyer-Lindquist coordinates. For a better visualization, we rescaled $F^{EM}_{\rm LNRF}$ by $r^2$. For high spin
runs ($a=0.9,\ 0.99,\ -0.9$), the presence of the negative Poynting flux (resembling negative energy states in the Penrose process) is quite prominent between the horizon and the ergosphere.

 An alternate way to understand the features in Fig.~\ref{fig:poynting_zamo} is through the membrane paradigm (\citealt{Thorne1986_book}). In the membrane paradigm (for details see Appendix \ref{sect:mem_para}), we can relate
the mass-energy entering/leaving the BH membrane ($dM_{\ast}/dt$) to the torque ($ dJ/dt$) exerted on the membrane and dissipation ($T_H dS_H/dt$) occurring on the membrane as
follows,
\be 
\label{eq:first_law}
\int_{r_H} \alpha  T^{\hat{r}}_{t_{BL}} \  dS^{\prime}_r = -\Omega_H \int_{r_H} \alpha T^{\hat{r}}_{\phi_{BL}} \  dS^{\prime}_r + \int_{r_H} \alpha^{2} T^{\hat{r}}_{\hat{t}} \  dS^{\prime}_r
\ee 
where $\alpha=\sqrt{\Sigma \Delta}/\sqrt{\Lambda}$ is the lapse function, with $\Sigma=r^2+a^2 \cos^2 \theta$, $\Delta= r^2+a^2 -2Mr$, $\Lambda=(r^2+a^2)^2 -a^2\Delta \sin^2 \theta$. $dS^{\prime}_r=\Lambda^{1/2} \sin \theta d\theta d\phi$ is the area element on the membrane. Here, $M_{\ast}$ and $J$ are the mass and angular momentum of the BH membrane, respectively. The angular velocity of the black hole is defined as $\Omega_H=a/2r_H$. Bekenstein-Hawking temperature and entropy are denoted by  $T_{H}$ and $S_H$ respectively.

The first and second terms on the RHS of equation (\ref{eq:first_law}) represent $\Omega_H dJ/dt$ and $T_H dS_H/dt$, respectively (for details see Appendix \ref{sect:mem_para}). According to the second law of BH thermodynamics, the third term in equation \ref{eq:first_law}, represents the `irreducible mass' of the BH, thermodynamically interpreted as stating that the entropy of the BH can never decrease (\citealt{Bardeen1973, Hawking1976}). Therefore, $T_H dS_H/dt >0$, hence $T^{\hat{r}}_{\hat{t}} > 0$.  We have also verified that both matter and EM parts of $T^{\hat{r}}_{\hat{t}}$ in our simulations follow this constraint (see Appendix \ref{sect:mem_para} and Fig. \ref{fig:ent_torw_dis}). 
Therefore, for all our runs, including the one for a non-spinning BH ($\Omega_H=0$) where the BZ mechanism is absent, we get a positive $T^{\hat{r}}_{\hat{t}}$ and hence a 
negative Poynting flux $F^{EM}_{\rm LNRF}=-\left(T^{\hat{r}}_{\hat{t}}\right)^{EM}$ at the horizon because of the mass-energy accretion.

\subsection{Magnetic flux threading the BH}
\label{sect:phi_bh_spin}
\begin{figure}
\centering
 \includegraphics[scale=0.48]{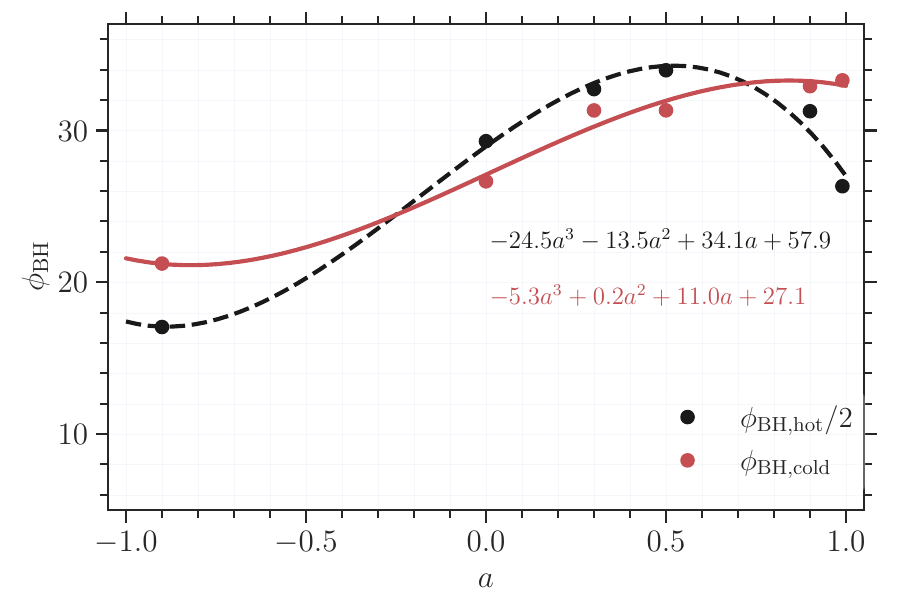}
\caption{Variation of the dimensionless magnetic flux parameter threading the event horizon with BH spin for a thin MAD (red solid line). For comparison, $\phi_{BH}(a)/2$ for the hot MAD in our simulation (black dashed line) is  also shown.}
\label{fig:phi_mad}
\end{figure}

It is evident from equation (\ref{eq:BZ_eff}) that the most important factor other than the BH spin in determining BZ efficiency is the amount of magnetic flux threading the BH horizon. Fig.~\ref{fig:phi_mad} shows that time-averaged (between $t=18000-20000$) values of $\phi_{BH}$ for the cold MAD for different spins. For comparison, we also show the time-averaged (between $t=8000-10000$)  profile of $\phi_{BH}(a)/2$  for the hot MAD in our simulation. We fit $\phi_{BH}(a)$ with a third-order polynomial giving rise to the best-fit values
\be
\label{eq:phi_bh_fit}
\begin{aligned}
\phi_{\rm BH, cold} &= -5.3 a^3 + 0.2a^2  + 11a + 27.1\\
 \phi_{\rm BH, hot} &= -24.5 a^3 - 13.5a^2  + 34.1a + 57.9
\end{aligned}
\ee
for cold and hot MADs, respectively.  It is interesting to note that although  $\phi_{BH}$ in our hot MAD simulations are slightly higher than that reported in  \cite{Narayan2022}, the overall spin-dependence is quite similar, suggesting that this is a robust feature independent of the different floors and initial conditions used in the current work and in \cite{Narayan2022}.

Similar to the hot MAD, we find that retrograde BHs saturate at a lower magnetic flux compared to their prograde counterpart in the thin MAD.  However, the value of the MAD parameter  $\phi_{\rm BH}(a)$ in the cold MAD is only about half the value of hot MAD's value in our simulations and has a much flatter profile. Interestingly, the peak value of $\phi_{BH}(a)$ occurs at a higher spin value of approximately $a \approx 0.9$ for thin cold MADs, in contrast to the lower spin value of $a=0.5$ observed in hot MADs.

\subsection{Energy extraction at the BH horizon}
\label{sect:em_energy_extraction}
\begin{figure}
\centering
 \includegraphics[scale=0.48]{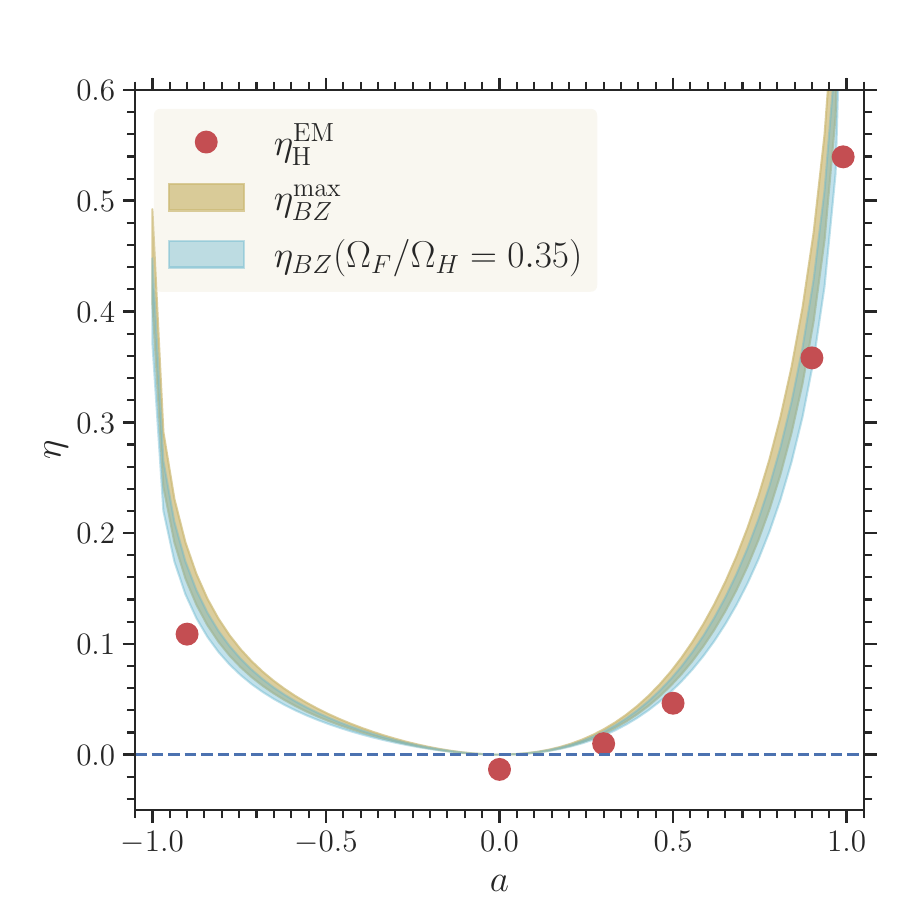 }
\caption{Variation of horizon electromagnetic efficiency $\eta^{EM}_H$ with black hole spin $a$ for the cold MAD. Each red-filled circle represent the time-averaged  (between $t=18000-20000$) value of $\eta^{EM}_H$ from a simulation. The yellow band represents the range of the maximum BZ efficiency $\eta^{\rm max}_{BZ}$ (with $\Omega_F/\Omega_H=0.5$) expected from spilt-monopole and paraboloidal magnetic fields. The cyan band illustrates the range of $\eta_{BZ}$ expected for spilt-monopole and paraboloidal magnetic fields for the value of $\Omega_F/\Omega_H=0.35$ obtained in our simulation. A close match between $\eta^{EM}_H$ obtained from simulation and BZ maximum efficiency indicates that the energy extraction process in our cold MAD simulations is the BZ process.}
\label{fig:eta_horizon}
\end{figure}

\begin{figure}
\centering
 \includegraphics[scale=0.5]{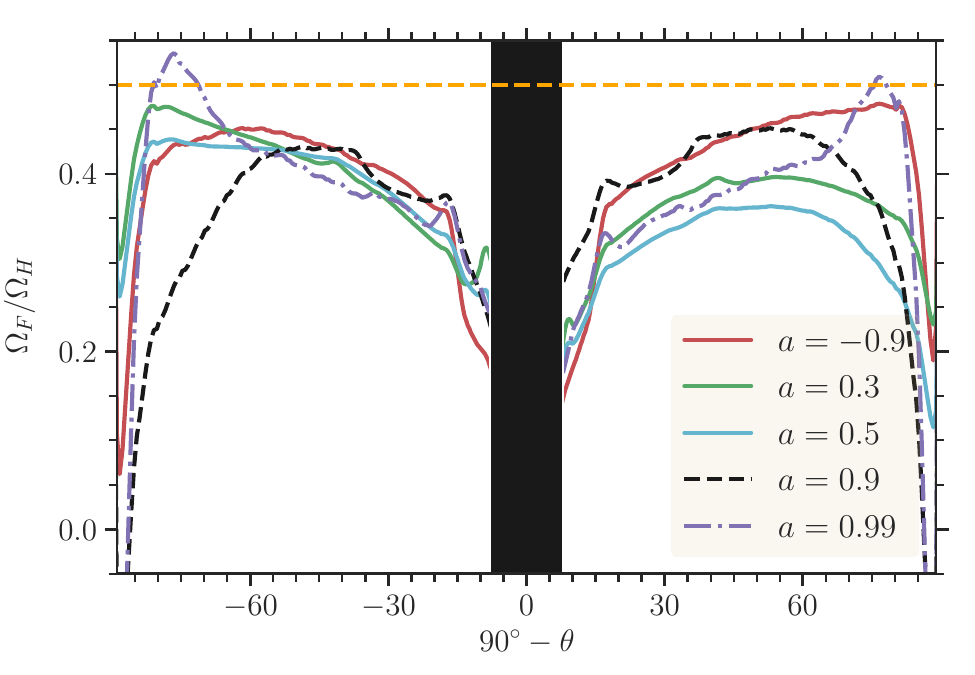}
\caption{Meridional profiles of normalized magnetic field line angular frequency $\Omega_F/\Omega_H$ at the event horizon for cold MADs; $\Omega_H$ is the angular frequency of the BH. The orange dashed line indicates the value of $\Omega_F/\Omega_H=0.5$ required for maximum BZ efficiency. A black mask is applied around the zero latitudes where force-free approximation fails. The value of $\Omega_F/\Omega_H$ increases with latitude, away from the disk, where the region becomes more force-free.  The ratio has an average value $\approx 0.35$, slightly lower than $0.5$.}
\label{fig:omega_f}
\end{figure}

\begin{figure}
\centering
 \includegraphics[scale=0.5]{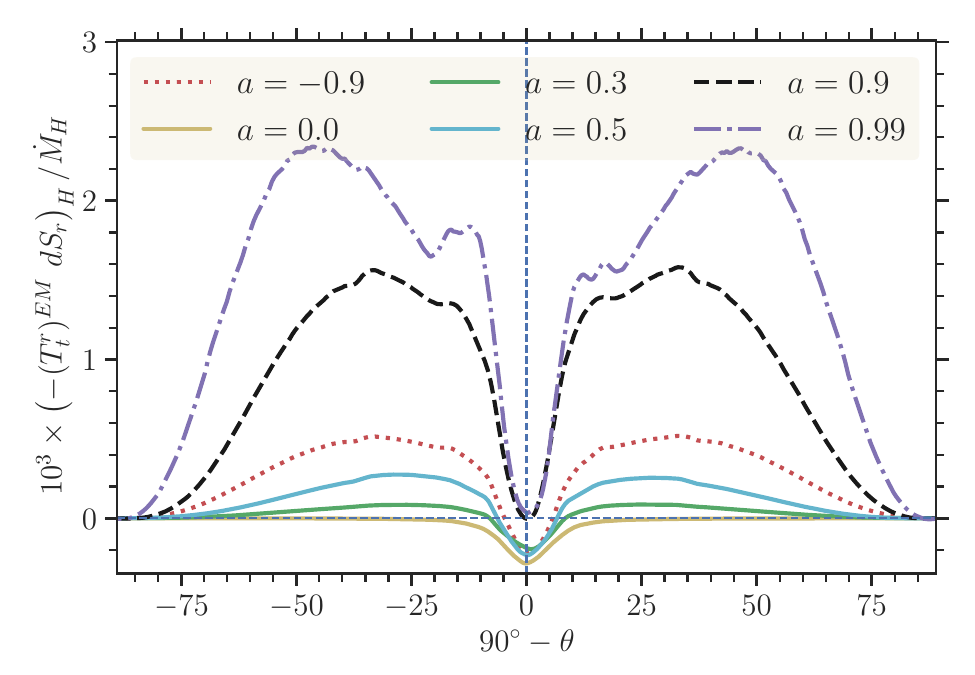}
\caption{Variation of area-weighted electromagnetic flux at the horizon with latitude $90^{\circ}-\theta$ for cold MADs with different BH spins. Most of the outflowing electromagnetic luminosity comes from the higher latitudes where force-free conditions prevail. }
\label{fig:T_rt_em_lat}
\end{figure}

In the previous section, we qualitatively showed the presence of an outward Poynting flux at the black hole's horizon in the coordinate frame. In this section, we show the rate of total electromagnetic energy going into/out of the event horizon for the cold MAD as a function of spin $a$. In  Fig.~\ref{fig:eta_horizon}, we normalize the energy inflow/outflow rate with the mass accretion rate at the horizon $\dot{M}_H$ to get the horizon electromagnetic efficiency $\eta^{EM}_H$ as defined in equation (\ref{eq:eta_em}). 

Red-filled circles in Fig.~\ref{fig:eta_horizon} represent the time-averaged (between $t=18000-20000$) $\eta^{EM}_H$ for each simulation of the cold MAD with different spin $a$. The yellow band shows a range of maximum BZ efficiencies $\eta^{\rm max}_{BZ}$ (obtained by assuming $\Omega_F/\Omega_H=0.5$ in equation \ref{eq:BZ_eff}) expected from the split-monopole and paraboloidal magnetic field geometries (for details see section \ref{sect:Bz_intro} and reference therein).  To calculate $\eta^{\rm max}_{BZ} (a)$ we use the fitted polynomial $\phi_{\rm BH, cold} (a)$ as shown in equation \ref{eq:phi_bh_fit}.   We find that a prograde BH around cold MAD
tends to have higher efficiency compared to its retrograde counterpart. However, the difference in efficiency between the prograde and retrograde BHs around a cold MAD turns out to be smaller than that found in the context of a hot MAD (see, e.g., Fig.~4 in \citealt{Narayan2022}). A weaker dependence of horizon electromagnetic efficiency on the sense of BH rotation in a cold MAD predominantly stems from the flatter profile of $\phi_{BH}(a)$ in the cold MAD compared to that in the hot MAD.

We find that $\eta^{EM}_H$ obtained from the simulations are slightly lower than the maximum BZ efficiency. The discrepancy mainly occurs for two apparent reasons: first, the value of $\Omega_F/\Omega_H$ is not equal to $0.5$ required for maximum BZ efficiency, and second, mass accretion impacts the BZ efficiency calculated in the force-free regime. We discuss both effects one by one.  

Fig.~\ref{fig:omega_f} shows the meridional profiles of magnetic field line frequency $\Omega_F$ (normalized by the  BH's angular frequency $\Omega_H$) at 
the event horizon for the cold MAD around spinning BHs. Out of different possible ways to calculate $\Omega_F$ (see, e.g., \citealt{Gammie2003}), we adopt the definition $\Omega_F=F_{t\theta}/F_{\theta \phi}$ which provides the least noisy profiles. We find that $\Omega_F/\Omega_H$ increases with latitude, away from the disk, where the region becomes more force-free; and again drops in the polar region. We suspect that the quick drop in  $\Omega_F/\Omega_H$ in the polar region
is likely due to the activation of floors in those regions. It is interesting to note that
the angle-averaged value of 
$(\Omega_F/\Omega_H)_{av} \approx 0.35$ is independent of the BH spin and slightly less than the 
value of $\Omega_F/\Omega_H = 0.5$ required for maximum BZ efficiency. This value $(\Omega_F/\Omega_H)_{av} \approx 0.35$ has also been found in earlier semi-thin/thin disk simulations (\citealt{Hawley2006, Penna2013}), albeit in weakly magnetized configurations, implying that the average value is likely to be independent of magnetic field strength in global MHD simulations of thin accretion disks. If we consider the modified  $\Omega_F/\Omega_H =  0.35$ in equation (\ref{eq:BZ_eff}), then the theoretical estimate of BZ efficiency as shown by the cyan band in Fig.~\ref{fig:eta_horizon} moves slightly  towards the red filled-circles, depicting the efficiency $\eta^{EM}_H$ directly calculated from the simulations of the cold MAD.

Next, we investigate the variation of normalized (by $\dot{M}_H$) electromagnetic flux $-(T^{r}_{t})^{EM}$ at the horizon with latitude ($=90^{\circ} - \theta$) with the following sign convention: positive sign implies outflow and vice versa (Fig.~\ref{fig:T_rt_em_lat}). We explicitly weighted $-(T^{r}_{t})^{EM}$ with area element
$dS_r(r=r_H)$ to account for contributions from different latitudes to $\eta^{EM}_H$. We see that
high latitude regions, where force-free conditions prevail, contribute most to $\eta^{EM}_H$, while the disk region, where mass accretion predominantly occurs, has an almost vanishing (for high prograde  spin BHs) or negative (for low prograde spin and retrograde BHs) contribution to the
electromagnetic horizon efficiency. This zero or negative contribution from the disk region to 
$\eta^{EM}_H$ further accounts for the discrepancy between the $\eta^{EM}_H$ and $\eta^{\rm max}_{BZ}$.

To summarize, while mass accretion has a minor effect on the horizon electromagnetic efficiency, $\eta^{EM}_H$, in cold MADs,  the overall efficiency is remarkably consistent with the expected electromagnetic efficiency predicted by the BZ mechanism.

\subsection{Jet efficiency}
\label{sect:results_eta_j}

\begin{figure}
\centering
 \includegraphics[scale=0.52]{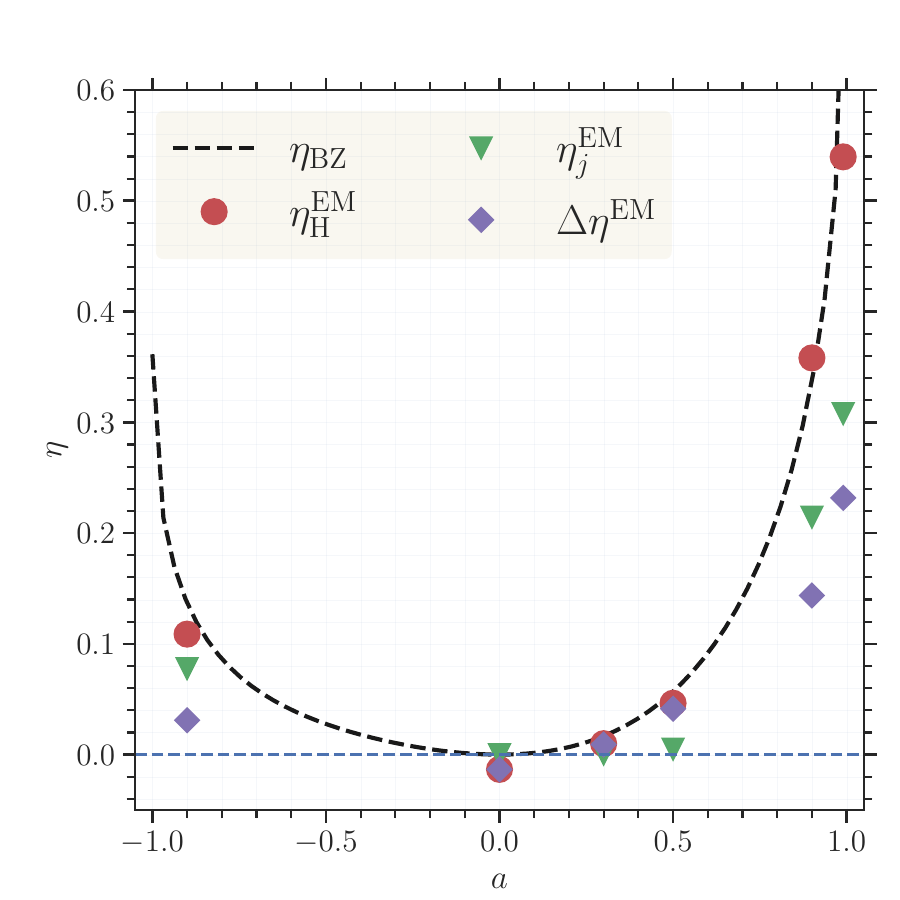}
\caption{Variation of electromagnetic jet efficiency $\eta^{EM}_j$ (for definition see section \ref{sect:diagnostics}), the
horizon efficiency $\eta^{EM}_H$ and their difference $\Delta \eta^{EM}=\eta^{EM}_H - \eta^{EM}_j$ with BH spin. The black dashed line represents the BZ efficiency with $\kappa=0.04$, which closely matches the horizon efficiency. Notably, the non-zero  $\Delta \eta^{EM}$  implies that only a fraction of the black hole's spin energy, extracted via the BZ mechanism, is actually used to power relativistic jets.}
\label{fig:eta_jet}
\end{figure}

The BZ process is quite effective in liberating energy, even in cold MADs. Nevertheless, it remains unclear whether the extracted energy is entirely channeled into the relativistic jet or is manifested in alternative forms, such as powering winds or disk heating. Although the design of our numerical experiments does not allow us to exactly quantify the fraction of BZ energy that goes into wind launching or is dissipated in the disk, it does allow us to measure the jet power. We measure the electromagnetic jet efficiency $\eta^{EM}_j$ by calculating the Poynting flux at $r=50$ in the relativistic polar regions (for details, see section \ref{sect:diagnostics}).

Fig.~\ref{fig:eta_jet} illustrates how the electromagnetic jet efficiency ($\eta^{EM}_j$) varies with black hole spin along with horizon efficiency ($\eta^{EM}_H$) and their difference ($\Delta \eta^{EM}=\eta^{EM}_H - \eta^{EM}_j$).  We see that jet efficiency $\eta^{EM}_j$ is much less than the horizon efficiency $\eta^{EM}_H$, indicating that electromagnetic energy 
extracted at the BH event horizon is not entirely channeled into the relativistic jets; only a fraction goes into powering the jets depending on the BH spin.
While highly spinning prograde black holes ($a=0.9, 0.99$) have a jet efficiency $\approx 20-30\%$, a high spinning retrograde ($a=-0.9$) BH has $\eta^{EM}_j\approx 8\%$. On the other hand, a low-spinning prograde ($a=0.5$) BH has a very low jet efficiency $\approx 1$ per cent. 

We compare our results with earlier studies of thin MADs with different black hole spins and disk thicknesses $H/R$. We quote the jet efficiencies for some representative studies of thin MADs: i) $\eta_{j}\approx 1\%$ for $H/R=0.1$ and $a=0.5$ (\citealt{Avara2016}), ii) $\eta_{j}\approx 50\%$ for $H/R=0.03$ and $a=0.9375$ (\citealt{Liska2022}), iii) $\eta_j \approx 10\%$ for $H/R=0.0375$ and $a=0.9375$ (\citealt{Scepi2024}).
We caution, however, that different studies use different definitions of `jet' and `jet efficiency'. For example, while \cite{Avara2016} and \cite{Scepi2024} calculate $\eta_j$  by calculating
Poynting flux at $r \approx 50$ $r_g$ in the highly magnetized polar regions ($\sigma \ge 1$), \cite{Liska2022} define $\eta_j$ as the normalized electromagnetic plus fluid energy fluxes at $r=5 \ r_g$ in the jet region.  Our definition of $\eta^{EM}_j$ (see section \ref{sect:diagnostics}) closely matches the definition of $\eta_j$ in \cite{Avara2016} and \cite{Scepi2024}, with the difference that we use a more conservative estimate of jet efficiency by considering only cells with relativistic flow in the polar region (for details see section \ref{sect:diagnostics}). 
On the other hand, we expect the horizon efficiency $\eta^{EM}_H$ in our work to match $\eta_j$ in \cite{Liska2022}, as energy flux in the highly magnetized polar region is predominantly due to electromagnetic flux.

Overall, inspection of previous studies and current work indicates that jet efficiency $\eta^{EM}_j$ is pretty small ($\approx 1\%$) for the slowly spinning BHs ($a<0.5$), but can be as large as $30\%$ for highly spinning prograde BHs, although only a fraction of the extracted BZ power goes into the jet.

\subsection{Radiative flux}
\label{sect:results_F_ff}

\begin{figure*}
\centering
 \includegraphics[scale=0.55]{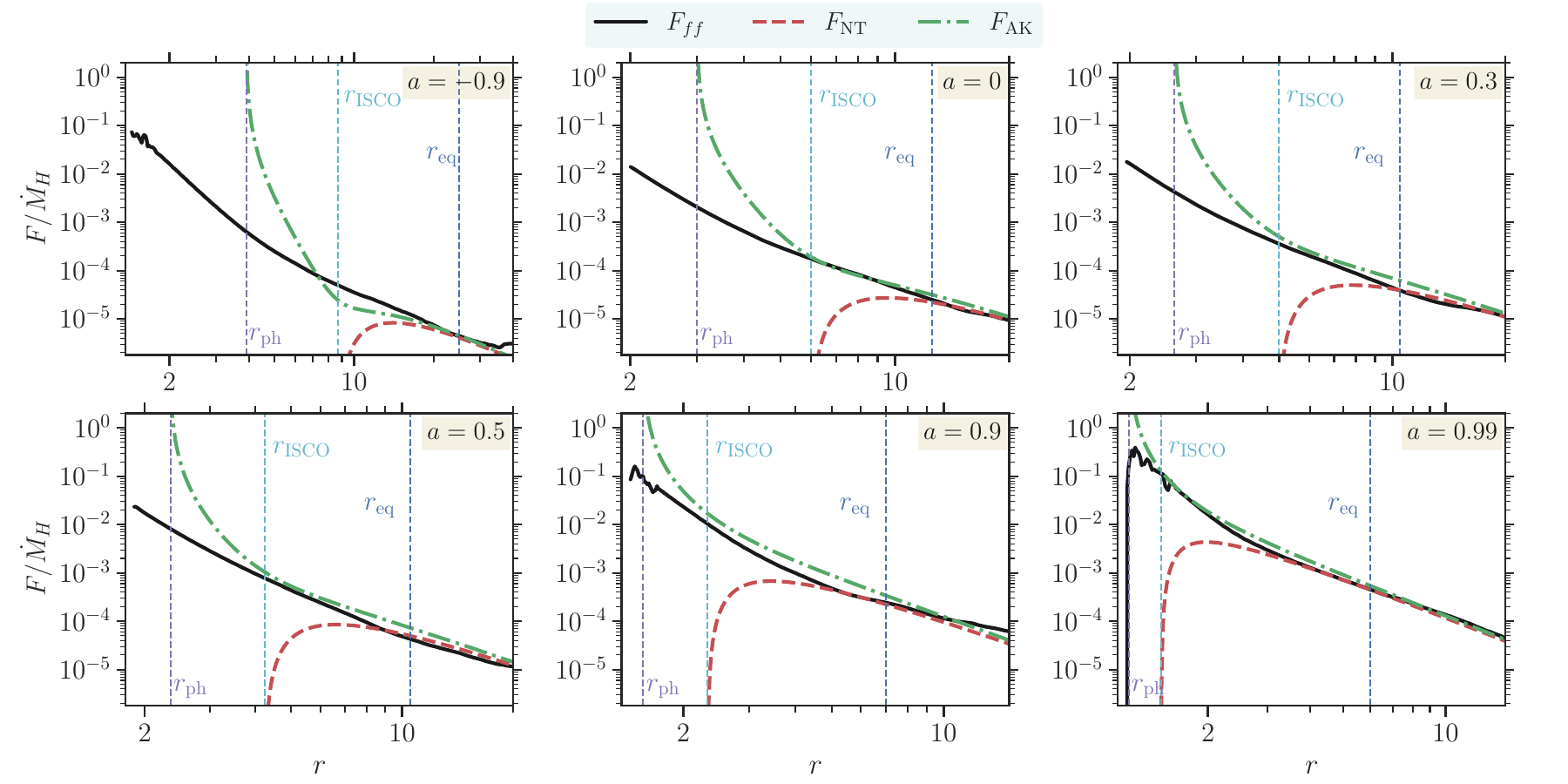}
\caption{Radial profiles of radiative flux $F_{ff}$ calculated in an orthonormal fluid frame for all the runs with different BH spin $a$. The profile of $F_{ff}(r)$ is compared to the radiative flux expected from a NT disk ($F_{NT}$) and from the AK model ($F_{AK}$). Vertical dashed lines 
indicate the photon orbit radius $r_{\rm ph}$ (magenta), ISCO $r_{\rm ISCO}$ (cyan) and inflow equilibrium radius $r_{\rm eq}$ (blue). }
\label{fig:rad_flux_ff}
\end{figure*}

% \begin{figure*}
% \centering
%  \includegraphics[scale=0.55]{stress_r_ff.pdf}
% \caption{Radial profiles of emissivity for different spin parameters.}
% \label{fig:wrphi_ff}
% \end{figure*}

The use of an optically thin cooling function to keep the disk thin allows us to directly measure the radiative efficiency of the disk. To compare with \cite{Novikov1973}, we measure the radiative flux $F_{ff}$ of the disk in an  orthonormal fluid frame (\citealt{Krolik2005, Penna2013, Dexter2016, White2019_tilt}) with a mean velocity $\bar{u}^{\mu}$ and defined by
\be
F_{ff} (r) = \frac{1}{\int dx^{\tilde{\phi}} (r,\theta=\pi/2) } \int \bar{\mathcal{S}} \ dx^{\tilde{\theta}} dx^{\tilde{\phi}}
\ee 
where $\bar{\mathcal{S}}$ is the time-averaged (between $t=18000-20000$) cooling function obtained from the simulation and $dx^{\tilde{\mu}}=e^{\tilde{\mu}}_{\nu} \ dx^{\nu}_{BL}$ is the
cell width in the fluid frame in the respective direction. Here
$dx^{\nu}_{BL} = \left[ \Delta t_{BL}, \  \Delta r, \  \Delta \theta, \ \Delta \phi_{BL} \right]$ is cell width in the Boyer-Lindquist coordinates
with $\Delta t_{BL}= -2Mr/(r^2 - 2Mr + a^2) \Delta r$ and $\Delta \phi_{BL}=\Delta \phi - a/(r^2 - 2Mr + a^2)\Delta r$. The integration is performed over all $\theta$ and $\phi$.

Fig.~\ref{fig:rad_flux_ff} shows the radial profiles of $F_{ff}$ for the cold MADs with different spin $a$. We scale $F_{ff}$ by the constant mass accretion rate (within the inflow equilibrium radius) as different simulations have different accretion rates. We also compare $F_{ff}$ with the radiative flux predicted from the NT ($F_{NT}$; \citealt{Novikov1973}) and AK ($F_{AK}$; \citealt{Agol_Krolik2000}) models. The calculation of $F_{AK}$ is sensitive to $\Delta \epsilon$, the constant defining the increment in radiative efficiency. We quantify $\Delta \epsilon$  by calculating the height-integrated (within $\pm 2H_{\rm th}$) fluid frame Maxwell stress $W^{\tilde{r}}_{\tilde{\phi}}$ defined by
\be
W^{\tilde{r}}_{\tilde{\phi}} (r) =  \frac{1}{\int dx^{\tilde{\phi}} (r,\theta=\pi/2) }
\int_{{\rm disk}} \left(T^{\tilde{r}}_{\tilde{\phi}} \right)^{EM}  \ dx^{\tilde{\theta}} dx^{\tilde{\phi}}
\ee 
at the ISCO (for details see, section 3.1 in \citealt{Beckwith2008}). Here, $u^{\tilde{\mu}} = e^{\tilde{\mu}}_{\nu} \ u^{\nu}_{BL}$ is the fluid frame four-velocity and  $\left(T^{\tilde{r}}_{\tilde{\phi}} \right)^{EM} =  b^2 u^{\tilde{r}} u_{\tilde{\phi}} - b^{\tilde{r}} b_{\tilde{\phi}}$ is the time-averaged
(between $t=18000-20000$) electromagnetic component of the stress-energy tensor in the fluid frame.

Fig.~\ref{fig:rad_flux_ff} clearly shows that the radiative flux $F_{ff}(r)$ in our simulation closely
matches $F_{NT}(r)$ in the outer disk within the inflow equilibrium radius $r_{eq}$, but differs significantly from it as the ISCO is approached. In addition, we observe the presence of substantial dissipation inside the ISCO, similar to what was observed in the earlier global simulations of weakly magnetized accretion flows (\citealt{Beckwith2008, Noble2009, Penna2010}) due to the presence of non-zero stress inside the ISCO. However, $F_{ff}(r)$ in our simulations is  only 66\% that in \cite{Scepi2024}.  We also observe that in our simulation, the contribution from the disk to the radiative flux is larger (around $75-80\%$) compared to that in \cite{Scepi2024} (around 66\%) .  We speculate that the difference is due to the difference in target scale height ($\epsilon_{\rm th}=0.1$ vs $\epsilon_{\rm th}=0.0375$) and different implementations of the cooling function.

The AK model predicts a slightly larger radiative flux $F_{AK}(r)$ compared to $F_{ff}$ and $F_{NT}$ for all of our simulations with prograde BH spins. In contrast, for the retrograde run ($a=-0.9$), $F_{AK}(r)$ is larger than  the NT flux at all radii but is less than $F_{ff}$ for $r>r_{\rm ISCO}$ and exceeds $F_{ff}$ in the
intra-ISCO region. The overestimate of dissipation in the AK model derives from the assumption that outward angular momentum transport in the disk is due to the turbulent radial Maxwell stress, leading to the dissipation of magnetic and kinetic energies into thermal energy (for a review see \citealt{Balbus1998}). However, in a thin MAD, the vertical Maxwell stress contributes significantly to
the angular momentum transport (wind-driven) and a fraction of the gravitational energy released is transported via winds (\citealt{Scepi2024}).

\subsection{Radiative efficiency}
\label{sect:results_eta_R}

\begin{figure}
\centering
 \includegraphics[scale=0.55]{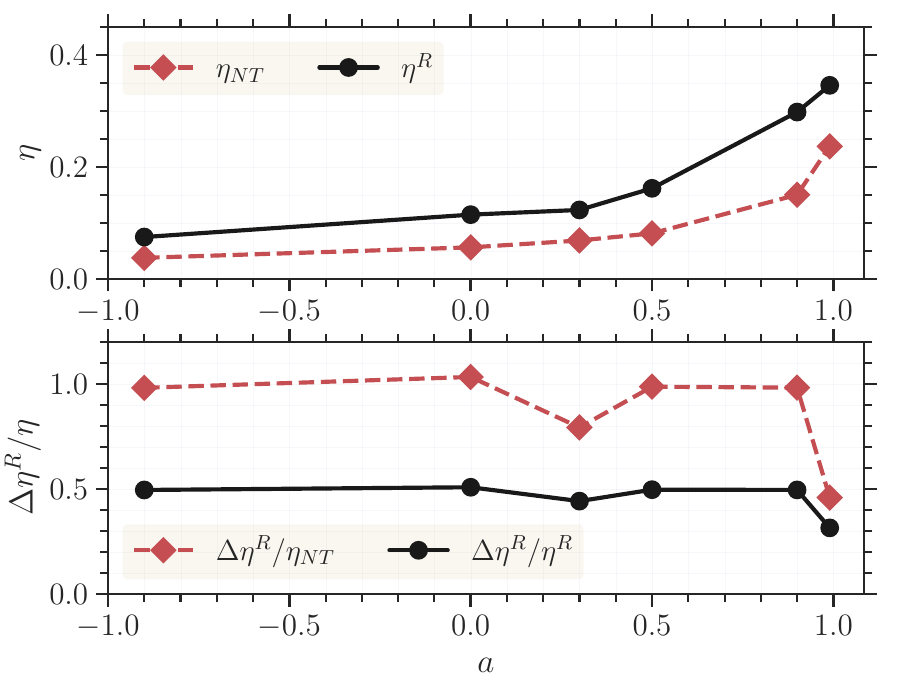}
\caption{Top panel: Variation of radiative efficiency $\eta^{R}$ with spin $a$ for the thin cold MAD. Spin dependence of modified Novikov-Thorne efficiency (for details see text) is also shown for comparison.  Bottom panel: Fractional change in radiative efficiency  $\Delta \eta^{R}/\eta_{NT}$ for the thin cold MAD with respect to NT disks. Thin cold MADs are approximately twice as radiatively efficient as  NT disks. }
\label{fig:eta_grtrans}
\end{figure}
 
The radiative efficiency $\eta^{R}$ can be calculated by radially integrating the radiative flux $F_{ff}(r)$ or equivalently by integrating the cooling term $\mathcal{S}_t$ over the four-volume $dt dV$ (see, e.g., \citealt{Avara2016} and Appendix \ref{sect:eta_appendix}).
It is crucial to consider that not all photons emitted by the disk will reach the observer at infinity, as some will be captured by the BH (\citealt{Thorne1974}), while others will be affected by gravitational redshift and beaming. To account for this, an inner integration limit $r_{\rm cap}$ is typically introduced, below which all emitted photons are assumed to be captured by the black hole.  Nevertheless, it has been found that the choice of $r_{\rm cap}$ significantly impacts the calculation of radiative efficiency for magnetized accretion flows (\citealt{Avara2016}; also see Appendix \ref{sect:eta_appendix}).  To circumvent these uncertainties, we employ ray tracing to calculate radiative efficiency using the radiative transfer code {\tt grtrans} (\citealt{Dexter2009, Dexter2016}) as described below.

To calculate $\eta^{R}$ for a BH of mass $M_{BH}=10 M_{\odot}$ and mass accretion rate $\dot{M} = 0.1 \dot{M}_{\rm Edd}$ for different spin values $a$, we use {\tt grtrans} with the following assumptions. We consider the disk to be optically thick and razor-thin, allowing us to describe its temperature using a blackbody temperature, given by 
$T_{\rm eff}=(F_{\rm fit}/\sigma_{SB})^{1/4}$, where $\sigma_{SB}$ is the Stephan-Boltzmann constant. The radiative flux of the disk, $F_{\rm fit}(r)$ is modeled by a broken power given by
\be
F_{\rm fit}(r) = c_0  \left[ c_1 \left(r/r_b \right)^{-\alpha_1}  e^ {-(r/r_b)} + c_2  \left(r/r_b \right)^{-\alpha_2} \right].
\ee 
This flux profile is equivalent to $F_{ff}$  within the inflow equilibrium radius $r_{eq}$ and transitions to $F_{NT}$ as it approaches $r_{eq}$. The coefficients $c_0$, $c_1$, $c_2$, $r_b$, $\alpha_1$ and $\alpha_2$ depend on spin $a$.  Additionally, we assume that the four-velocity of the disk is the same as that of a standard Novikov-Thorne (NT) disk, based on the expectation that in the region close to the BH, where radial velocity is significantly large, gravitational redshift dominates beaming effects that arise due to rapid infall of plasma toward the BH horizon.
 
We positioned ten cameras, each with a size of $600 \ r_g  \times 600 \ r_g$ and resolution $\Delta \alpha=\Delta \beta=0.6 \ r_g$, at equal intervals in terms of the cosine of the inclination angle $i \in [2.75^{\circ}, 87.5^{\circ}]$, at an infinite distance, to compute the solid-angle averaged radiative luminosity $L_{\nu}$ ($\nu$ is the frequency of the observed radiation) given by 
\be 
 L_{\nu} = 4 \pi \Delta \alpha \Delta \beta \sum_{i} \sum_{\rm pixels} I_{\nu} ({\rm pixels}) \ d(\cos i).
\ee
Finally, radiative efficiency is calculated as $\eta^{R} = L_{\rm bol}/\dot{M}c^2$, where $L_{\rm bol}=\int L_{\nu} \ d\nu$ is the bolometric luminosity. Here we note that the presence of a non-zero effective temperature within the intra-ISCO region causes a shift in the peak of the spectrum to higher frequencies, a phenomenon that has been previously observed and reported in earlier studies (\citealt{Zhu2012, Hankla2022, Mummery2024}).

The top panel of Fig.~\ref{fig:eta_grtrans} illustrates how the radiative efficiency $\eta^{R}$ changes with the spin of the black hole for a thin cold MAD. For comparison, we also show the NT efficiency computed using ray tracing methods. The resulting values, represented by the red dashed lines in the top panel of Fig. \ref{fig:eta_grtrans}, exhibit slight deviations from the standard NT efficiency \citep{Novikov1973}, particularly for high-spin prograde black holes.  This discrepancy is consistent with the findings of \cite{Thorne1974}, who considered photon capture by the BH in their calculation.
The bottom panel illustrates the fractional change in radiative efficiency $\Delta \eta^{R}/\eta_{NT}$ between thin cold MADs and NT disks. We also show the fractional change in $\Delta \eta^{R}$ with respect to $\eta^{R}$, whose significance is discussed in section \ref{sect:discuss_corona}.
Notably, thin cold MADs exhibit a significant increase in luminosity, by a factor of $50-100\%$, depending on the spin, compared to their NT counterparts, a finding consistent with the $30-100$ per cent enhancement calculated for a moderately magnetized thin disk of thickness $\epsilon_{\rm th}=0.05$ by \cite{Kinch2021}. In contrast, weakly magnetized (SANE) thin disks show a more modest enhancement of only $5-10\%$ (\citealt{Penna2013}). Interestingly, the fractional change in $\eta^{R}$ remains relatively constant with respect to black hole spin $a$. This finding is consistent with previous studies of thin MAD, which reported similar enhancements (around 80\%) for thin cold MADs with different scale-heights and black hole spins: $\epsilon_{\rm th}=0.1, \ a=0.5$ (\citealt{Avara2016}), and  $\epsilon_{\rm th}=0.0375, \ a=0.9375$ (\citealt{Scepi2024}), respectively. 

\section{Discussion}
\label{sect:discussion}

\subsection{Weaker spin dependence of MAD parameter in the thin MAD}
\label{sect:discuss_phi_bh}
Magnetic field flux threading the BH, $\phi_{BH}$, is one of the primary factors in determining the MAD-ness of the accretion flow and hence the efficiency of the BZ process. In the context of hot accretion flows, a prograde BH is found to possess a larger magnetic flux threading the event horizon compared to a retrograde BH, leading to more powerful jets for prograde BHs with the same spin parameter 
$|a|$ (\citealt{Tchekhovskoy2010, McKinney2012, Narayan2022}). However, some analytical studies suggest a contrasting scenario, where a retrograde BH surrounded by a geometrically thin disk, with an enlarged plunging region, is predicted to be threaded by a stronger magnetic flux due to enhanced magnetic flux trapping, potentially yielding more powerful jets (\citealt{Reynolds2006, Garofalo2009}).

Our simulations of thin MADs with thermal scale-height $H_{\rm th}/R\approx 0.1$  reveal a similar trend of magnetic flux threading the black hole ($\phi_{BH}$) with spin parameter $a$, but with two notable differences. Firstly, the dependence of $\phi_{BH}$ on spin $a$ is weaker in thin MADs than in hot MADs. Secondly, the maximum value of $\phi_{BH}$ is attained at a higher spin $a\approx 0.9$ for thin MADs, whereas in hot MADs, the maximum occurs around $a\approx 0.5$. We hypothesize that these differences arise from the interaction between the disk and black hole spin, where $\phi_{BH}$ is maximized when the black hole's angular velocity ($\Omega_H$) resonates with the disk's angular velocity ($\Omega$). In contrast to hot MADs, where the accretion flow is sub-Keplerian ($\Omega \approx 0.5 \Omega_K$), thin MADs exhibit Keplerian rotation (see Fig.~\ref{fig:h_r_omega_a9}). As a result, the peak in the $\phi_{BH}-a$ relation shifts towards higher $\Omega_H$	values, corresponding to higher spin parameters $a$. Additionally, from a geometrical perspective, the impact of the accretion flow on magnetic field saturation at the BH horizon is more pronounced in geometrically thick hot MADs ($H/R \approx 0.3-0.5$) compared to geometrically thin MADs with $H_{\rm th}/R\approx 0.1$. However, we caution that a more rigorous investigation is necessary to fully understand the long-standing issue of flux saturation in the near-horizon region.

\subsection{Plausible manifestation of BZ power}
\label{sect:discuss_bz}
Our simulations of thin MADs around spinning BHs have successfully demonstrated the presence of the Blandford-Znajek (BZ) mechanism \citep{Blandford1977}. We calculated the horizon electromagnetic efficiency, $\eta^{EM}_H$, which measures the BZ power extracted from the BH's rotational energy. 
Our findings indicate that only a portion of this power is channeled into the jet, while the remaining power is potentially used to launch winds or dissipate in the disk. However, distinguishing the contribution of BZ power to winds and radiation is difficult to disentangle. This is because winds (\citealt{Blandford1982}) and dissipation (\citealt{Brandenburg1995, Parkin2014b}) are also powered by accretion energy. In a radiatively efficient thin disk, cooling balances the heating. However, the radiative efficiency $\eta^{R}$, which measures the radiative power in our simulations, does not capture the total energy dissipation in the disk. Additional cooling mechanisms, including adiabatic cooling due to fluid expansion and advective cooling caused by accretion onto the black hole (\citealt{Ichimaru1977, Rees1982}), can also significantly contribute to cooling the disk. As a result, the complex interplay between these processes makes it challenging to determine the relative contributions of BZ power to winds and disk heating.

However, we do find that geometrically thin accretion disks in the  MAD configuration are radiatively $50-100\%$ more efficient than the NT disk. Although this excess luminosity $\Delta \eta^{R} \dot{M} c^2$ is primarily due to enhanced magnetic dissipation in the thin MAD (see \citealt{Scepi2024} for detailed discussion), we suspect that it  can have a contribution from remaining BZ power that is not channeled into the jets.

\subsection{Astrophysical implications of hot plunging region in a thin MAD}
\label{sect:discuss_corona}
A substantial fraction of the excess luminosity, $\Delta \eta^{R} \dot{M} c^2$, originates from the intra-ISCO region (as seen in Fig.~\ref{fig:rad_flux_ff}) in the highly radiatively efficient thin magnetically arrested disk (MAD). Previous studies of weakly magnetized disks (e.g., \citealt{Zhu2012}) found that intra-ISCO emission can impact black hole spin measurements based on continuum methods such as {\tt KERBB} (\citealt{Li2005}) and {\tt BHSPEC} (\citealt{Davis2005}), which assume no emission from the plunging region following the NT model.  We suspect that this effect will be more pronounced in thin MADs, where the intra-ISCO region is significantly hotter (as effective temperature $\propto F_{ff}^{1/4}$) than in a weakly magnetized disk.

Not all intra-ISCO emission contributes to thermal blackbody radiation; a fraction of it can produce a power-law dominated spectrum due to thermal electrons (in weakly magnetized flows; \citealt{Zhu2012}) or non-thermal electrons (in highly magnetized accretion flows; \citealt{Hankla2022}). Although our single-temperature GRMHD simulation of a thin MAD does not account for the role of electrons in producing spectra, by assuming a fraction $\delta_{\rm cor}$ of the excess luminosity forms a corona with coronal luminosity $L_{\rm cor}= \delta_{\rm cor} \ \Delta \eta^{R} \dot{m} c^2$, we can correlate $\Delta \eta^{R}/\eta^{R}$ (see the bottom panel of Fig. \ref{fig:eta_grtrans}) to
$L_{\rm cor}/(L_{\rm cor} + L_{\rm disk})$ as follows:
\be
\label{eq:L_cor}
%\begin{aligned}
 \frac{L_{\rm cor}}{L_{\rm cor} + L_{\rm disk}} = \delta_{\rm cor} \left( \frac{\Delta \eta^{R}}{\eta^{R}} \right),
%\end{aligned}
\ee
where we assume the disk luminosity is $L_{\rm disk}= \eta^{R} \dot{m} c^2 - L_{\rm cor}$. Note that the coronal luminosity $L_{\rm cor}$ can also receive contributions from the hot, optically thin upper layer of the disk, leading to spectral hardening (\citealt{Jiang2014b, Scepi2024}).

Additionally, numerous AGN transients (\citealt{Trakhtenbrot2019, Ricci2020}) exhibit rapid UV/X-ray evolution, suggesting changes in coronal dissipation or jet activity. These events are characterized by rapid timescale evolution and spectral changes. A large amount of energy dissipation close to the BH surrounded by a strongly magnetized thin disk (\citealt{Dexter2019, Scepi2021}) could potentially contribute to powering such and similar events in these sources.

\subsection{Jets in the high soft state of XRBs}
Our simulations suggest that thin MADs exhibit moderate efficiency in extracting the rotational energy of the black hole (BH), in contrast to hot MADs, which display higher efficiency (\citealt{Tchekhovskoy2011}). A significant fraction of the BZ power is channeled into powerful Poynting flux outflows in the polar regions, which are often associated with jet emission. These outflows are observed as kinetic power during state transitions and radio flux in the hard state (\citealt{Fender2004, Remillard2006}). However, in the soft state, typically characterized by a thin disk (\citealt{Done2007}), observational upper limits on radio flux are often reported. The lack of observed radio flux in the soft state may be attributed to the lower values of MAD parameter $\phi_{BH}$ associated with our simulated jets, compared to those in strongly magnetized hot accretion flows, or to less efficient conversion of Poynting flux to radio emission in these jets.

Alternatively, it is also possible that the geometrically thin disk may be unable to sustain large-scale magnetic fields, which are necessary for jet production, over a sufficient period. This is because the dynamo process in a thin disk is efficient in generating alternate polarity fields (\citealt{Gressel2010, Flock2012a, Dhang2024}), which may weaken the pre-existing large-scale field that is inherited from the hot flows in the truncated disk during the low-hard state (\citealt{Begelman2014, Dhang2020}). Our current simulations, which spanned $10^4 \ r_g/c$ in the cold phase, equivalent to just $0.5$ seconds for a $10 \ M_{\odot}$ BH accreting source, are not long enough to fully address this issue, and we reserve it for future investigation.

\section{Summary}
\label{sect:summary}
We performed a set of GRMHD simulations to systematically study highly magnetized geometrically thin disks (thin MADs) around BHs of different spins, focusing primarily on the extraction of energy from the accretion flow and the BH spin. Here we list the significant findings in our work.
\begin{itemize}
    \item We have characterized the properties of a thin, cold MAD around a BH, finding that it is typified by a MAD parameter of $\phi_{BH} \approx 30$,  a highly magnetized disk body with $\beta \sim 1$, regular but small-amplitude eruptions, and a moderately high energy extraction efficiency (including both matter and electromagnetic components) at the horizon (see section \ref{sect:madness}).
    
    \item For a given BH spin, we find that the MAD parameter $\phi_{BH}$ in thin, cold MADs is approximately half the value of its hot counterpart and exhibits a weaker dependence on spin compared to hot MADs (Sections \ref{sect:phi_bh_spin} and \ref{sect:discuss_phi_bh}).
    
    \item We have explicitly demonstrated the presence of the Blandford-Znajek (BZ) mechanism in thin MADs surrounding spinning BHs (see Section \ref{sect:Bz_intro}). Our results show that only a fraction ( $10 \%-70 \%$ ) of the extracted BZ power ($\eta^{EM}_H \dot{M} c^2$), is channeled into the jet ($\eta^{EM}_j \dot{M}c^2$), with the remaining energy ($\Delta \eta^{EM}\dot{M} c^2$) potentially used to launch winds or power radiation from the disk/corona (see sections \ref{sect:em_energy_extraction}, \ref{sect:results_eta_j}, and \ref{sect:discuss_bz}).
    
    \item Similar to earlier studies, we find that thin MADs are highly radiatively efficient, with efficiencies $50-100\%$ higher than a standard disk, depending on the BH spin (see section \ref{sect:results_eta_R}). We attribute this excess luminosity primarily to the enhanced magnetic dissipation in the intra-ISCO region along with a fraction of BZ power that is not used to power the jet and connect it to the formation of coronae in accreting sources (see section \ref{sect:discuss_corona}).
\end{itemize}

We acknowledge support from NSF grant AST 1303335, NASA Astrophysics Theory Program grants NNX17AK55G and 80NSSC22K0826, NSF AST 230798380, NASA NSSC24K1094 and TM4-25005X, and an Alfred P. Sloan Fellowship (JD). PD thanks Chris White and Xuening Bai for numerous discussions on the {\tt Athena++} code and implementation of the cooling function which was developed for a different project. We also thank Feryal Ozel, Chris Done, Julian Krolik, George Wong and the anonymous referee for the useful comments and suggestions. All the computations were performed in NASA's HECC facility (Pleiades) and CU Boulder's Alpine Cluster.  Visualizations are done using VisIt (\citealt{HPV:VisIt}) and Matplotlib (\citealt{Hunter:2007}).

\bibliography{bibtex}{}

\begin{thebibliography}{}
\expandafter\ifx\csname natexlab\endcsname\relax\def\natexlab#1{#1}\fi
\providecommand{\url}[1]{\href{#1}{#1}}
\providecommand{\dodoi}[1]{doi:~\href{http://doi.org/#1}{\nolinkurl{#1}}}
\providecommand{\doeprint}[1]{\href{http://ascl.net/#1}{\nolinkurl{http://ascl.net/#1}}}
\providecommand{\doarXiv}[1]{\href{https://arxiv.org/abs/#1}{\nolinkurl{https://arxiv.org/abs/#1}}}

\bibitem[{{Agol} \& {Krolik}(2000)}]{Agol_Krolik2000}
{Agol}, E., \& {Krolik}, J.~H. 2000, \apj, 528, 161, \dodoi{10.1086/308177}

\bibitem[{{Avara} {et~al.}(2016){Avara}, {McKinney}, \& {Reynolds}}]{Avara2016}
{Avara}, M.~J., {McKinney}, J.~C., \& {Reynolds}, C.~S. 2016, \mnras, 462, 636,
  \dodoi{10.1093/mnras/stw1643}

\bibitem[{Bai \& Stone(2013)}]{Bai2013}
Bai, X.-N., \& Stone, J.~M. 2013, \apj, 769, 76,
  \dodoi{10.1088/0004-637X/769/1/76}

\bibitem[{Balbus \& Hawley(1991)}]{Balbus1991}
Balbus, S.~A., \& Hawley, J.~F. 1991, \apj, 376, 214, \dodoi{10.1086/170270}

\bibitem[{{Balbus} \& {Hawley}(1998)}]{Balbus1998}
{Balbus}, S.~A., \& {Hawley}, J.~F. 1998, Reviews of Modern Physics, 70, 1,
  \dodoi{10.1103/RevModPhys.70.1}

\bibitem[{{Bardeen} {et~al.}(1973){Bardeen}, {Carter}, \&
  {Hawking}}]{Bardeen1973}
{Bardeen}, J.~M., {Carter}, B., \& {Hawking}, S.~W. 1973, Communications in
  Mathematical Physics, 31, 161, \dodoi{10.1007/BF01645742}

\bibitem[{{Bardeen} {et~al.}(1972){Bardeen}, {Press}, \&
  {Teukolsky}}]{Bardeen1972}
{Bardeen}, J.~M., {Press}, W.~H., \& {Teukolsky}, S.~A. 1972, \apj, 178, 347,
  \dodoi{10.1086/151796}

\bibitem[{{Beckwith} {et~al.}(2008){Beckwith}, {Hawley}, \&
  {Krolik}}]{Beckwith2008}
{Beckwith}, K., {Hawley}, J.~F., \& {Krolik}, J.~H. 2008, \apj, 678, 1180,
  \dodoi{10.1086/533492}

\bibitem[{Begelman \& Armitage(2014)}]{Begelman2014}
Begelman, M.~C., \& Armitage, P.~J. 2014, \apjl, 782, L18,
  \dodoi{10.1088/2041-8205/782/2/L18}

\bibitem[{{Begelman} \& {Armitage}(2023)}]{Begelman2023}
{Begelman}, M.~C., \& {Armitage}, P.~J. 2023, \mnras, 521, 5952,
  \dodoi{10.1093/mnras/stad914}

\bibitem[{{Begelman} \& {Pringle}(2007)}]{Begelman_Pringle2007}
{Begelman}, M.~C., \& {Pringle}, J.~E. 2007, \mnras, 375, 1070,
  \dodoi{10.1111/j.1365-2966.2006.11372.x}

\bibitem[{{Begelman} {et~al.}(2022){Begelman}, {Scepi}, \&
  {Dexter}}]{Begelman2022}
{Begelman}, M.~C., {Scepi}, N., \& {Dexter}, J. 2022, \mnras, 511, 2040,
  \dodoi{10.1093/mnras/stab3790}

\bibitem[{Bisnovatyi-Kogan \& Ruzmaikin(1974)}]{Bisnovatyi-Kogan1974a}
Bisnovatyi-Kogan, G.~S., \& Ruzmaikin, A.~A. 1974, \apss, 28, 45,
  \dodoi{10.1007/BF00642237}

\bibitem[{{Blandford} \& {Begelman}(1999)}]{Blandford_Begelman1999}
{Blandford}, R.~D., \& {Begelman}, M.~C. 1999, \mnras, 303, L1,
  \dodoi{10.1046/j.1365-8711.1999.02358.x}

\bibitem[{Blandford \& Payne(1982)}]{Blandford1982}
Blandford, R.~D., \& Payne, D.~G. 1982, \mnras, 199, 883,
  \dodoi{10.1093/mnras/199.4.883}

\bibitem[{Blandford \& Znajek(1977)}]{Blandford1977}
Blandford, R.~D., \& Znajek, R.~L. 1977, \mnras, 179, 433,
  \dodoi{10.1093/mnras/179.3.433}

\bibitem[{Brandenburg {et~al.}(1995)Brandenburg, Nordlund, Stein, \&
  Torkelsson}]{Brandenburg1995}
Brandenburg, A., Nordlund, A., Stein, R.~F., \& Torkelsson, U. 1995, \apj, 446,
  741, \dodoi{10.1086/175831}

\bibitem[{{Cao}(2011)}]{Cao2011}
{Cao}, X. 2011, \apj, 737, 94, \dodoi{10.1088/0004-637X/737/2/94}

\bibitem[{{Chakrabarti}(1989)}]{Chakrabarti1989}
{Chakrabarti}, S.~K. 1989, \apj, 347, 365, \dodoi{10.1086/168125}

\bibitem[{Chandrasekhar(1960)}]{Chandrasekhar1960a}
Chandrasekhar, S. 1960, Proceedings of the National Academy of Science, 46,
  253, \dodoi{10.1073/pnas.46.2.253}

\bibitem[{Childs {et~al.}(2012)Childs, Brugger, Whitlock, Meredith, Ahern,
  Pugmire, Biagas, Miller, Harrison, Weber, Krishnan, Fogal, Sanderson, Garth,
  Bethel, Camp, R\"{u}bel, Durant, Favre, \& Navr\'{a}til}]{HPV:VisIt}
Childs, H., Brugger, E., Whitlock, B., {et~al.} 2012, in High Performance
  Visualization--Enabling Extreme-Scale Scientific Insight, 357--372,
  \dodoi{10.1201/b12985}

\bibitem[{Colella \& Woodward(1984)}]{Colella1984}
Colella, P., \& Woodward, P.~R. 1984, Journal of Computational Physics, 54,
  174, \dodoi{10.1016/0021-9991(84)90143-8}

\bibitem[{{Davis} {et~al.}(2005){Davis}, {Blaes}, {Hubeny}, \&
  {Turner}}]{Davis2005}
{Davis}, S.~W., {Blaes}, O.~M., {Hubeny}, I., \& {Turner}, N.~J. 2005, \apj,
  621, 372, \dodoi{10.1086/427278}

\bibitem[{{Dexter}(2016)}]{Dexter2016}
{Dexter}, J. 2016, \mnras, 462, 115, \dodoi{10.1093/mnras/stw1526}

\bibitem[{{Dexter} \& {Agol}(2009)}]{Dexter2009}
{Dexter}, J., \& {Agol}, E. 2009, \apj, 696, 1616,
  \dodoi{10.1088/0004-637X/696/2/1616}

\bibitem[{{Dexter} \& {Begelman}(2019)}]{Dexter2019}
{Dexter}, J., \& {Begelman}, M.~C. 2019, \mnras, 483, L17,
  \dodoi{10.1093/mnrasl/sly213}

\bibitem[{{Dhang} {et~al.}(2023){Dhang}, {Bai}, \& {White}}]{Dhang2023}
{Dhang}, P., {Bai}, X.-N., \& {White}, C.~J. 2023, \apj, 944, 182,
  \dodoi{10.3847/1538-4357/acb534}

\bibitem[{{Dhang} {et~al.}(2020){Dhang}, {Bendre}, {Sharma}, \&
  {Subramanian}}]{Dhang2020}
{Dhang}, P., {Bendre}, A., {Sharma}, P., \& {Subramanian}, K. 2020, \mnras,
  494, 4854, \dodoi{10.1093/mnras/staa996}

\bibitem[{{Dhang} {et~al.}(2024){Dhang}, {Bendre}, \&
  {Subramanian}}]{Dhang2024}
{Dhang}, P., {Bendre}, A.~B., \& {Subramanian}, K. 2024, \mnras, 530, 2778,
  \dodoi{10.1093/mnras/stae1011}

\bibitem[{Dhang \& Sharma(2019)}]{Dhang2019}
Dhang, P., \& Sharma, P. 2019, \mnras, 482, 848, \dodoi{10.1093/mnras/sty2692}

\bibitem[{Done {et~al.}(2007)Done, Gierli{\'n}ski, \& Kubota}]{Done2007}
Done, C., Gierli{\'n}ski, M., \& Kubota, A. 2007, \aapr, 15, 1,
  \dodoi{10.1007/s00159-007-0006-1}

\bibitem[{Einfeldt(1988)}]{Einfeldt1988}
Einfeldt, B. 1988, 671.
\newblock \url{https://ui.adsabs.harvard.edu/abs/1988stw..proc..671E}

\bibitem[{{Event Horizon Telescope Collaboration} {et~al.}(2019){Event Horizon
  Telescope Collaboration}, {Akiyama}, {Alberdi}, {Alef}, {Asada}, {Azulay},
  {Baczko}, {Ball}, {Balokovi{\'c}}, {Barrett}, {Bintley}, {Blackburn},
  {Boland}, {Bouman}, {Bower}, {Bremer}, {Brinkerink}, {Brissenden}, {Britzen},
  {Broderick}, {Broguiere}, {Bronzwaer}, {Byun}, {Carlstrom}, {Chael}, {Chan},
  {Chatterjee}, {Chatterjee}, {Chen}, {Chen}, {Cho}, {Christian}, {Conway},
  {Cordes}, {Crew}, {Cui}, {Davelaar}, {De Laurentis}, {Deane}, {Dempsey},
  {Desvignes}, {Dexter}, {Doeleman}, {Eatough}, {Falcke}, {Fish}, {Fomalont},
  {Fraga-Encinas}, {Friberg}, {Fromm}, {G{\'o}mez}, {Galison}, {Gammie},
  {Garc{\'\i}a}, {Gentaz}, {Georgiev}, {Goddi}, {Gold}, {Gu}, {Gurwell},
  {Hada}, {Hecht}, {Hesper}, {Ho}, {Ho}, {Honma}, {Huang}, {Huang}, {Hughes},
  {Ikeda}, {Inoue}, {Issaoun}, {James}, {Jannuzi}, {Janssen}, {Jeter}, {Jiang},
  {Johnson}, {Jorstad}, {Jung}, {Karami}, {Karuppusamy}, {Kawashima},
  {Keating}, {Kettenis}, {Kim}, {Kim}, {Kim}, {Kino}, {Koay}, {Koch}, {Koyama},
  {Kramer}, {Kramer}, {Krichbaum}, {Kuo}, {Lauer}, {Lee}, {Li}, {Li},
  {Lindqvist}, {Liu}, {Liuzzo}, {Lo}, {Lobanov}, {Loinard}, {Lonsdale}, {Lu},
  {MacDonald}, {Mao}, {Markoff}, {Marrone}, {Marscher}, {Mart{\'\i}-Vidal},
  {Matsushita}, {Matthews}, {Medeiros}, {Menten}, {Mizuno}, {Mizuno}, {Moran},
  {Moriyama}, {Moscibrodzka}, {Mul{\ensuremath{\ddot{}}}ler}, {Nagai}, {Nagar},
  {Nakamura}, {Narayan}, {Narayanan}, {Natarajan}, {Neri}, {Ni}, {Noutsos},
  {Okino}, {Olivares}, {Oyama}, {{\"O}zel}, {Palumbo}, {Patel}, {Pen}, {Pesce},
  {Pi{\'e}tu}, {Plambeck}, {PopStefanija}, {Porth}, {Prather},
  {Preciado-L{\'o}pez}, {Psaltis}, {Pu}, {Ramakrishnan}, {Rao}, {Rawlings},
  {Raymond}, {Rezzolla}, {Ripperda}, {Roelofs}, {Rogers}, {Ros}, {Rose},
  {Roshanineshat}, {Rottmann}, {Roy}, {Ruszczyk}, {Ryan}, {Rygl},
  {S{\'a}nchez}, {S{\'a}nchez-Arguelles}, {Sasada}, {Savolainen}, {Schloerb},
  {Schuster}, {Shao}, {Shen}, {Small}, {Sohn}, {SooHoo}, {Tazaki}, {Tiede},
  {Tilanus}, {Titus}, {Toma}, {Torne}, {Trent}, {Trippe}, {Tsuda}, {van
  Bemmel}, {van Langevelde}, {van Rossum}, {Wagner}, {Wardle}, {Weintroub},
  {Wex}, {Wharton}, {Wielgus}, {Wong}, {Wu}, {Young}, {Young}, {Younsi},
  {Yuan}, {Yuan}, {Zensus}, {Zhao}, {Zhao}, {Zhu}, {Anczarski}, {Baganoff},
  {Eckart}, {Farah}, {Haggard}, {Meyer-Zhao}, {Michalik}, {Nadolski},
  {Neilsen}, {Nishioka}, {Nowak}, {Pradel}, {Primiani}, {Souccar},
  {Vertatschitsch}, {Yamaguchi}, \& {Zhang}}]{EHT_V_2019}
{Event Horizon Telescope Collaboration}, {Akiyama}, K., {Alberdi}, A., {et~al.}
  2019, \apjl, 875, L5, \dodoi{10.3847/2041-8213/ab0f43}

\bibitem[{{Fender} {et~al.}(2004){Fender}, {Belloni}, \& {Gallo}}]{Fender2004}
{Fender}, R.~P., {Belloni}, T.~M., \& {Gallo}, E. 2004, \mnras, 355, 1105,
  \dodoi{10.1111/j.1365-2966.2004.08384.x}

\bibitem[{{Fishbone} \& {Moncrief}(1976)}]{Fishbone1976}
{Fishbone}, L.~G., \& {Moncrief}, V. 1976, \apj, 207, 962,
  \dodoi{10.1086/154565}

\bibitem[{Flock {et~al.}(2012)Flock, Dzyurkevich, Klahr, Turner, \&
  Henning}]{Flock2012a}
Flock, M., Dzyurkevich, N., Klahr, H., Turner, N., \& Henning, T. 2012, \apj,
  744, 144, \dodoi{10.1088/0004-637X/744/2/144}

\bibitem[{Fragile {et~al.}(2012)Fragile, Wilson, \& Rodriguez}]{Fragile2012a}
Fragile, P.~C., Wilson, J., \& Rodriguez, M. 2012, \mnras, 424, 524,
  \dodoi{10.1111/j.1365-2966.2012.21222.x}

\bibitem[{{Gammie} {et~al.}(2003){Gammie}, {McKinney}, \&
  {T{\'o}th}}]{Gammie2003}
{Gammie}, C.~F., {McKinney}, J.~C., \& {T{\'o}th}, G. 2003, \apj, 589, 444,
  \dodoi{10.1086/374594}

\bibitem[{{Gardiner} \& {Stone}(2005)}]{Gardiner2005}
{Gardiner}, T.~A., \& {Stone}, J.~M. 2005, Journal of Computational Physics,
  205, 509, \dodoi{10.1016/j.jcp.2004.11.016}

\bibitem[{{Garofalo}(2009)}]{Garofalo2009}
{Garofalo}, D. 2009, \apj, 699, 400, \dodoi{10.1088/0004-637X/699/1/400}

\bibitem[{Gressel(2010)}]{Gressel2010}
Gressel, O. 2010, \mnras, 405, 41, \dodoi{10.1111/j.1365-2966.2010.16440.x}

\bibitem[{Gressel \& Pessah(2015)}]{Gressel2015}
Gressel, O., \& Pessah, M.~E. 2015, \apj, 810, 59,
  \dodoi{10.1088/0004-637X/810/1/59}

\bibitem[{Guilet \& Ogilvie(2012)}]{Guilet2012}
Guilet, J., \& Ogilvie, G.~I. 2012, \mnras, 424, 2097,
  \dodoi{10.1111/j.1365-2966.2012.21361.x}

\bibitem[{{Hagen} \& {Done}(2023)}]{Hagen2023}
{Hagen}, S., \& {Done}, C. 2023, \mnras, 525, 3455,
  \dodoi{10.1093/mnras/stad2499}

\bibitem[{{Hankla} {et~al.}(2022){Hankla}, {Scepi}, \& {Dexter}}]{Hankla2022}
{Hankla}, A.~M., {Scepi}, N., \& {Dexter}, J. 2022, \mnras, 515, 775,
  \dodoi{10.1093/mnras/stac1785}

\bibitem[{{Hawking}(1976)}]{Hawking1976}
{Hawking}, S.~W. 1976, \prd, 13, 191, \dodoi{10.1103/PhysRevD.13.191}

\bibitem[{{Hawley} \& {Krolik}(2006)}]{Hawley2006}
{Hawley}, J.~F., \& {Krolik}, J.~H. 2006, \apj, 641, 103,
  \dodoi{10.1086/500385}

\bibitem[{Hawley {et~al.}(2013)Hawley, Richers, Guan, \& Krolik}]{Hawley2013}
Hawley, J.~F., Richers, S.~A., Guan, X., \& Krolik, J.~H. 2013, \apj, 772, 102,
  \dodoi{10.1088/0004-637X/772/2/102}

\bibitem[{Hunter(2007)}]{Hunter:2007}
Hunter, J.~D. 2007, Computing in Science \& Engineering, 9, 90,
  \dodoi{10.1109/MCSE.2007.55}

\bibitem[{{Ichimaru}(1977)}]{Ichimaru1977}
{Ichimaru}, S. 1977, \apj, 214, 840, \dodoi{10.1086/155314}

\bibitem[{Igumenshchev {et~al.}(2003)Igumenshchev, Narayan, \&
  Abramowicz}]{Igumenshchev2003}
Igumenshchev, I.~V., Narayan, R., \& Abramowicz, M.~A. 2003, \apj, 592, 1042,
  \dodoi{10.1086/375769}

\bibitem[{{Jacquemin-Ide} {et~al.}(2024){Jacquemin-Ide}, {Rincon},
  {Tchekhovskoy}, \& {Liska}}]{Jacquemin-Ide2024}
{Jacquemin-Ide}, J., {Rincon}, F., {Tchekhovskoy}, A., \& {Liska}, M. 2024,
  \mnras, 532, 1522, \dodoi{10.1093/mnras/stae1538}

\bibitem[{{Jiang} {et~al.}(2019){Jiang}, {Blaes}, {Stone}, \&
  {Davis}}]{Jiang2019}
{Jiang}, Y.-F., {Blaes}, O., {Stone}, J.~M., \& {Davis}, S.~W. 2019, \apj, 885,
  144, \dodoi{10.3847/1538-4357/ab4a00}

\bibitem[{Jiang {et~al.}(2014)Jiang, Stone, \& Davis}]{Jiang2014b}
Jiang, Y.-F., Stone, J.~M., \& Davis, S.~W. 2014, \apj, 784, 169,
  \dodoi{10.1088/0004-637X/784/2/169}

\bibitem[{{Kinch} {et~al.}(2021){Kinch}, {Schnittman}, {Noble}, {Kallman}, \&
  {Krolik}}]{Kinch2021}
{Kinch}, B.~E., {Schnittman}, J.~D., {Noble}, S.~C., {Kallman}, T.~R., \&
  {Krolik}, J.~H. 2021, \apj, 922, 270, \dodoi{10.3847/1538-4357/ac2b9a}

\bibitem[{{Krolik} {et~al.}(2005){Krolik}, {Hawley}, \& {Hirose}}]{Krolik2005}
{Krolik}, J.~H., {Hawley}, J.~F., \& {Hirose}, S. 2005, \apj, 622, 1008,
  \dodoi{10.1086/427932}

\bibitem[{{Li} {et~al.}(2005){Li}, {Zimmerman}, {Narayan}, \&
  {McClintock}}]{Li2005}
{Li}, L.-X., {Zimmerman}, E.~R., {Narayan}, R., \& {McClintock}, J.~E. 2005,
  \apjs, 157, 335, \dodoi{10.1086/428089}

\bibitem[{{Liska} {et~al.}(2022){Liska}, {Musoke}, {Tchekhovskoy}, {Porth}, \&
  {Beloborodov}}]{Liska2022}
{Liska}, M.~T.~P., {Musoke}, G., {Tchekhovskoy}, A., {Porth}, O., \&
  {Beloborodov}, A.~M. 2022, \apjl, 935, L1, \dodoi{10.3847/2041-8213/ac84db}

\bibitem[{Lubow {et~al.}(1994)Lubow, Papaloizou, \& Pringle}]{Lubow1994}
Lubow, S.~H., Papaloizou, J. C.~B., \& Pringle, J.~E. 1994, \mnras, 267, 235,
  \dodoi{10.1093/mnras/267.2.235}

\bibitem[{McKinney {et~al.}(2012)McKinney, Tchekhovskoy, \&
  Blandford}]{McKinney2012}
McKinney, J.~C., Tchekhovskoy, A., \& Blandford, R.~D. 2012, \mnras, 423, 3083,
  \dodoi{10.1111/j.1365-2966.2012.21074.x}

\bibitem[{{Miller} \& {Stone}(2000)}]{Miller2000}
{Miller}, K.~A., \& {Stone}, J.~M. 2000, \apj, 534, 398, \dodoi{10.1086/308736}

\bibitem[{{Mishra} {et~al.}(2020){Mishra}, {Begelman}, {Armitage}, \&
  {Simon}}]{Mishra2020}
{Mishra}, B., {Begelman}, M.~C., {Armitage}, P.~J., \& {Simon}, J.~B. 2020,
  \mnras, 492, 1855, \dodoi{10.1093/mnras/stz3572}

\bibitem[{{Motta} {et~al.}(2009){Motta}, {Belloni}, \& {Homan}}]{Motta2009}
{Motta}, S., {Belloni}, T., \& {Homan}, J. 2009, \mnras, 400, 1603,
  \dodoi{10.1111/j.1365-2966.2009.15566.x}

\bibitem[{{Mummery} {et~al.}(2024){Mummery}, {Ingram}, {Davis}, \&
  {Fabian}}]{Mummery2024}
{Mummery}, A., {Ingram}, A., {Davis}, S., \& {Fabian}, A. 2024, \mnras, 531,
  366, \dodoi{10.1093/mnras/stae1160}

\bibitem[{{Narayan} {et~al.}(2022){Narayan}, {Chael}, {Chatterjee}, {Ricarte},
  \& {Curd}}]{Narayan2022}
{Narayan}, R., {Chael}, A., {Chatterjee}, K., {Ricarte}, A., \& {Curd}, B.
  2022, \mnras, 511, 3795, \dodoi{10.1093/mnras/stac285}

\bibitem[{Narayan {et~al.}(2003)Narayan, Igumenshchev, \&
  Abramowicz}]{Narayan2003}
Narayan, R., Igumenshchev, I.~V., \& Abramowicz, M.~A. 2003, \pasj, 55, L69,
  \dodoi{10.1093/pasj/55.6.L69}

\bibitem[{Narayan {et~al.}(2012)Narayan, S{\"A dowski}, Penna, \&
  Kulkarni}]{Narayan2012}
Narayan, R., S{\"A dowski}, A., Penna, R.~F., \& Kulkarni, A.~K. 2012, \mnras,
  426, 3241, \dodoi{10.1111/j.1365-2966.2012.22002.x}

\bibitem[{Narayan \& Yi(1994)}]{Narayan1994}
Narayan, R., \& Yi, I. 1994, \apjl, 428, L13, \dodoi{10.1086/187381}

\bibitem[{{Nathanail} {et~al.}(2022){Nathanail}, {Dhang}, \&
  {Fromm}}]{Nathanail2022}
{Nathanail}, A., {Dhang}, P., \& {Fromm}, C.~M. 2022, \mnras, 513, 5204,
  \dodoi{10.1093/mnras/stac1276}

\bibitem[{Nemmen {et~al.}(2014)Nemmen, Storchi-Bergmann, \&
  Eracleous}]{Nemmen2014}
Nemmen, R.~S., Storchi-Bergmann, T., \& Eracleous, M. 2014, \mnras, 438, 2804,
  \dodoi{10.1093/mnras/stt2388}

\bibitem[{{Noble} {et~al.}(2009){Noble}, {Krolik}, \& {Hawley}}]{Noble2009}
{Noble}, S.~C., {Krolik}, J.~H., \& {Hawley}, J.~F. 2009, \apj, 692, 411,
  \dodoi{10.1088/0004-637X/692/1/411}

\bibitem[{{Novikov} \& {Thorne}(1973)}]{Novikov1973}
{Novikov}, I.~D., \& {Thorne}, K.~S. 1973, in Black Holes (Les Astres Occlus),
  343--450

\bibitem[{Parkin(2014)}]{Parkin2014b}
Parkin, E.~R. 2014, \mnras, 441, 2078, \dodoi{10.1093/mnras/stu699}

\bibitem[{{Penna} {et~al.}(2013){Penna}, {Kulkarni}, \&
  {Narayan}}]{Penna2013_init}
{Penna}, R.~F., {Kulkarni}, A., \& {Narayan}, R. 2013, \aap, 559, A116,
  \dodoi{10.1051/0004-6361/201219666}

\bibitem[{{Penna} {et~al.}(2010){Penna}, {McKinney}, {Narayan}, {Tchekhovskoy},
  {Shafee}, \& {McClintock}}]{Penna2010}
{Penna}, R.~F., {McKinney}, J.~C., {Narayan}, R., {et~al.} 2010, \mnras, 408,
  752, \dodoi{10.1111/j.1365-2966.2010.17170.x}

\bibitem[{Penna {et~al.}(2013)Penna, Narayan, \& S{\c a}dowski}]{Penna2013}
Penna, R.~F., Narayan, R., \& S{\c a}dowski, A. 2013, \mnras, 436, 3741,
  \dodoi{10.1093/mnras/stt1860}

\bibitem[{Penrose(1969)}]{Penrose1969}
Penrose, R. 1969, Riv. Nuovo Cim., 1, 252, \dodoi{10.1023/A:1016578408204}

\bibitem[{{Pessah}(2010)}]{Pessah2010}
{Pessah}, M.~E. 2010, \apj, 716, 1012, \dodoi{10.1088/0004-637X/716/2/1012}

\bibitem[{{Porth} {et~al.}(2021){Porth}, {Mizuno}, {Younsi}, \&
  {Fromm}}]{Porth2021}
{Porth}, O., {Mizuno}, Y., {Younsi}, Z., \& {Fromm}, C.~M. 2021, \mnras, 502,
  2023, \dodoi{10.1093/mnras/stab163}

\bibitem[{{Rees} {et~al.}(1982){Rees}, {Begelman}, {Blandford}, \&
  {Phinney}}]{Rees1982}
{Rees}, M.~J., {Begelman}, M.~C., {Blandford}, R.~D., \& {Phinney}, E.~S. 1982,
  \nat, 295, 17, \dodoi{10.1038/295017a0}

\bibitem[{Remillard \& McClintock(2006)}]{Remillard2006}
Remillard, R.~A., \& McClintock, J.~E. 2006, \araa, 44, 49,
  \dodoi{10.1146/annurev.astro.44.051905.092532}

\bibitem[{{Reynolds} {et~al.}(2006){Reynolds}, {Garofalo}, \&
  {Begelman}}]{Reynolds2006}
{Reynolds}, C.~S., {Garofalo}, D., \& {Begelman}, M.~C. 2006, \apj, 651, 1023,
  \dodoi{10.1086/507691}

\bibitem[{{Ricci} {et~al.}(2020){Ricci}, {Kara}, {Loewenstein}, {Trakhtenbrot},
  {Arcavi}, {Remillard}, {Fabian}, {Gendreau}, {Arzoumanian}, {Li}, {Ho},
  {MacLeod}, {Cackett}, {Altamirano}, {Gandhi}, {Kosec}, {Pasham}, {Steiner},
  \& {Chan}}]{Ricci2020}
{Ricci}, C., {Kara}, E., {Loewenstein}, M., {et~al.} 2020, \apjl, 898, L1,
  \dodoi{10.3847/2041-8213/ab91a1}

\bibitem[{{Ripperda} {et~al.}(2022){Ripperda}, {Liska}, {Chatterjee}, {Musoke},
  {Philippov}, {Markoff}, {Tchekhovskoy}, \& {Younsi}}]{Ripperda2022}
{Ripperda}, B., {Liska}, M., {Chatterjee}, K., {et~al.} 2022, \apjl, 924, L32,
  \dodoi{10.3847/2041-8213/ac46a1}

\bibitem[{{Sadowski}(2016)}]{Sadowski2016}
{Sadowski}, A. 2016, \mnras, 462, 960, \dodoi{10.1093/mnras/stw1852}

\bibitem[{{Salas} {et~al.}(2024){Salas}, {Musoke}, {Chatterjee}, {Markoff},
  {Porth}, {Liska}, \& {Ripperda}}]{Salas2024}
{Salas}, L.~D.~S., {Musoke}, G., {Chatterjee}, K., {et~al.} 2024, \mnras, 533,
  254, \dodoi{10.1093/mnras/stae1834}

\bibitem[{{Salvesen} {et~al.}(2016){Salvesen}, {Simon}, {Armitage}, \&
  {Begelman}}]{Salvesen2016}
{Salvesen}, G., {Simon}, J.~B., {Armitage}, P.~J., \& {Begelman}, M.~C. 2016,
  \mnras, 457, 857, \dodoi{10.1093/mnras/stw029}

\bibitem[{{Scepi} {et~al.}(2021){Scepi}, {Begelman}, \& {Dexter}}]{Scepi2021}
{Scepi}, N., {Begelman}, M.~C., \& {Dexter}, J. 2021, \mnras, 502, L50,
  \dodoi{10.1093/mnrasl/slab002}

\bibitem[{{Scepi} {et~al.}(2024){Scepi}, {Begelman}, \& {Dexter}}]{Scepi2024}
---. 2024, \mnras, 527, 1424, \dodoi{10.1093/mnras/stad3299}

\bibitem[{{Scepi} {et~al.}(2022){Scepi}, {Dexter}, \& {Begelman}}]{Scepi2022}
{Scepi}, N., {Dexter}, J., \& {Begelman}, M.~C. 2022, \mnras, 511, 3536,
  \dodoi{10.1093/mnras/stac337}

\bibitem[{Shakura \& Sunyaev(1973)}]{Shakura1973}
Shakura, N.~I., \& Sunyaev, R.~A. 1973, \aap, 24, 337.
\newblock \url{https://ui.adsabs.harvard.edu/abs/1973A%26A....24..337S}

\bibitem[{Sorathia {et~al.}(2012)Sorathia, Reynolds, Stone, \&
  Beckwith}]{Sorathia2012}
Sorathia, K.~A., Reynolds, C.~S., Stone, J.~M., \& Beckwith, K. 2012, \apj,
  749, 189, \dodoi{10.1088/0004-637X/749/2/189}

\bibitem[{{Stone} {et~al.}(2020){Stone}, {Tomida}, {White}, \&
  {Felker}}]{Stone2020}
{Stone}, J.~M., {Tomida}, K., {White}, C.~J., \& {Felker}, K.~G. 2020, \apjs,
  249, 4, \dodoi{10.3847/1538-4365/ab929b}

\bibitem[{{Tchekhovskoy} {et~al.}(2010){Tchekhovskoy}, {Narayan}, \&
  {McKinney}}]{Tchekhovskoy2010}
{Tchekhovskoy}, A., {Narayan}, R., \& {McKinney}, J.~C. 2010, \apj, 711, 50,
  \dodoi{10.1088/0004-637X/711/1/50}

\bibitem[{Tchekhovskoy {et~al.}(2011)Tchekhovskoy, Narayan, \&
  McKinney}]{Tchekhovskoy2011}
Tchekhovskoy, A., Narayan, R., \& McKinney, J.~C. 2011, \mnras, 418, L79,
  \dodoi{10.1111/j.1745-3933.2011.01147.x}

\bibitem[{{Thorne}(1974)}]{Thorne1974}
{Thorne}, K.~S. 1974, \apj, 191, 507, \dodoi{10.1086/152991}

\bibitem[{{Thorne} {et~al.}(1986){Thorne}, {Price}, \&
  {MacDonald}}]{Thorne1986_book}
{Thorne}, K.~S., {Price}, R.~H., \& {MacDonald}, D.~A. 1986, {Black holes: The
  membrane paradigm}

\bibitem[{{Trakhtenbrot} {et~al.}(2019){Trakhtenbrot}, {Arcavi}, {MacLeod},
  {Ricci}, {Kara}, {Graham}, {Stern}, {Harrison}, {Burke}, {Hiramatsu},
  {Hosseinzadeh}, {Howell}, {Smartt}, {Rest}, {Prieto}, {Shappee}, {Holoien},
  {Bersier}, {Filippenko}, {Brink}, {Zheng}, {Li}, {Remillard}, \&
  {Loewenstein}}]{Trakhtenbrot2019}
{Trakhtenbrot}, B., {Arcavi}, I., {MacLeod}, C.~L., {et~al.} 2019, \apj, 883,
  94, \dodoi{10.3847/1538-4357/ab39e4}

\bibitem[{Velikhov(1959)}]{Velikhov1959}
Velikhov, E. 1959, Sov. Phys. JETP, 36, 995

\bibitem[{{White} {et~al.}(2019{\natexlab{a}}){White}, {Quataert}, \&
  {Blaes}}]{White2019_tilt}
{White}, C.~J., {Quataert}, E., \& {Blaes}, O. 2019{\natexlab{a}}, \apj, 878,
  51, \dodoi{10.3847/1538-4357/ab089e}

\bibitem[{{White} {et~al.}(2020){White}, {Quataert}, \& {Gammie}}]{White2020}
{White}, C.~J., {Quataert}, E., \& {Gammie}, C.~F. 2020, \apj, 891, 63,
  \dodoi{10.3847/1538-4357/ab718e}

\bibitem[{{White} {et~al.}(2016){White}, {Stone}, \& {Gammie}}]{White2016}
{White}, C.~J., {Stone}, J.~M., \& {Gammie}, C.~F. 2016, \apjs, 225, 22,
  \dodoi{10.3847/0067-0049/225/2/22}

\bibitem[{{White} {et~al.}(2019{\natexlab{b}}){White}, {Stone}, \&
  {Quataert}}]{White_mad_2019}
{White}, C.~J., {Stone}, J.~M., \& {Quataert}, E. 2019{\natexlab{b}}, \apj,
  874, 168, \dodoi{10.3847/1538-4357/ab0c0c}

\bibitem[{{Zhu} {et~al.}(2012){Zhu}, {Davis}, {Narayan}, {Kulkarni}, {Penna},
  \& {McClintock}}]{Zhu2012}
{Zhu}, Y., {Davis}, S.~W., {Narayan}, R., {et~al.} 2012, \mnras, 424, 2504,
  \dodoi{10.1111/j.1365-2966.2012.21181.x}

\bibitem[{{Znajek}(1977)}]{Znajek1977}
{Znajek}, R.~L. 1977, \mnras, 179, 457, \dodoi{10.1093/mnras/179.3.457}

\end{thebibliography}
\bibliographystyle{aasjournal}

\appendix 
\label{sect:appen}
\restartappendixnumbering

\section{Quality factors and magnetic tilt angle in our simulations}
\label{sect:convergence_appendix}
\begin{figure*}
\centering
 \includegraphics[scale=0.6]{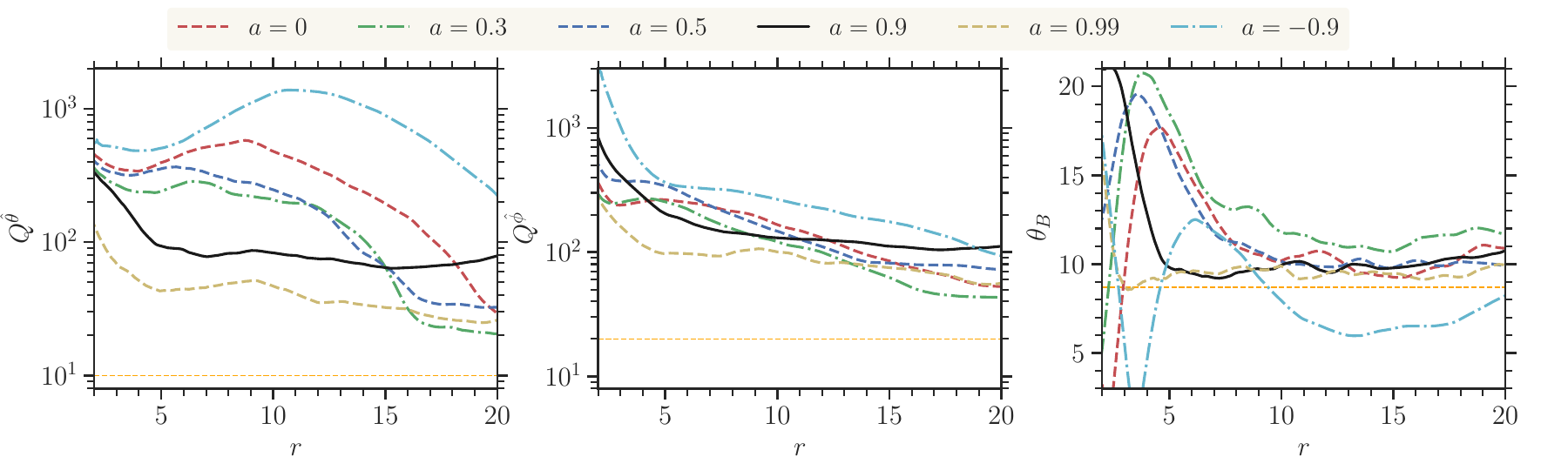}
\caption{Radial profiles of average quality factors $Q_{\theta}$, $Q_{\phi}$ and magnetic tilt angle $\theta_B$ for all the runs. The average is done over all $\phi$ and within $\pm 2H_{th}$ in the meridional direction. The yellow dashed line in each panel denotes the critical values of the convergence metrics prescribed by previous studies. }
\label{fig:convergence}
\end{figure*}

 Convergence studies are essential for establishing the credibility of numerical simulations. In ideal MHD simulations, without explicit dissipation, the concept of convergence is ill-defined. This is because, in ideal MHD, there are no fixed viscous and resistive length scales, and the dissipation scales depend on the grid size. Therefore, as the resolution is increased, new structures are created. Due to the lack of explicit dissipation, convergence in ideal MHD is typically demonstrated by varying the resolution and verifying that physical observables, such as the mass accretion rate, become insensitive to resolution above a certain threshold.

GRMHD simulations are computationally expensive, and running simulations with increasing resolution for convergence studies can sometimes become impractical. However, previous studies of magnetized accretion flows have found that a few numerical metrics such as quality factors
\be
 Q_{\theta} = \frac{2 \pi}{\Omega} \frac{| \bar{b}^{\tilde{\theta}} |} {\sqrt{{\overline{w}_{\rm tot}}}} \frac{1} {dx^{\tilde{\theta}}}, \\ \ \ 
 Q_{\phi} =  \frac{2 \pi}{\Omega} \frac{| \bar{b}^{\tilde{\phi}} |} {\sqrt{{\overline{w}_{\rm tot}}}}  \frac{1} {dx^{\tilde{\phi}}}
\label{eq:quality_f}
\ee 
and magnetic tilt angle, 
\be
\theta_{B} = \frac{1}{2} \sin^{-1} \alpha_{\rm mag} =  -\frac{1}{2} \sin^{-1} \left(\frac{\overline{b^{\tilde{r}}b^{\tilde{\phi}}}} {\overline{p}_{\rm mag}} \right)
\label{eq:tilt_f}
\ee
where $w_{\rm tot} (r,\theta) = \rho h + 2p_{\rm mag}$ is the total enthalpy, are useful in assessing convergence in MRI simulations (e.g., \citealt{Sorathia2012, Hawley2013, Dhang2019}). Although the quality factors indicating the number of grid points per critical wavelength in the respective direction were originally used to investigate the resolvability of MRI, they were found to be useful even in the non-linear regime of MRI-driven turbulent accretion flow (\citealt{Hawley2013}). On the other hand, the value of the magnetic tilt angle above a critical value ensures MRI saturation (\citealt{Pessah2010}).

Notably, these studies have primarily focused on weakly magnetized disks. Furthermore, in MADs, the fastest-growing mode of MRI $\lambda_{\rm MRI} = 2 \pi|b^{\tilde{\theta}}|/\Omega \sqrt{w_{\rm tot}}$ is larger than the scale height $H$, which complicates the interpretation of quality factors (\citealt{Begelman2022}). Despite these caveats, we employ quality factors and the magnetic tilt angle to compare the resolution of our simulations with those of previous MAD simulations.

Fig. \ref{fig:convergence} illustrates the radial profiles of average quality factors $Q_{\theta}$, $Q_{\phi}$ and magnetic tilt angle $\theta_B$ for all the runs. The quality factors are large at all radii as expected in a MAD (\citealt{McKinney2012, White_mad_2019, Salas2024}) and exceed the recommended values of $Q_{\theta,{\rm crit}}=10$ and $Q_{\phi,{\rm crit}}=20$ (yellow dashed line in each panel), suggested by earlier studies (see, e.g., \citealt{Hawley2013}). However, while the magnetic tilt angle is above the critical value $\theta_{B, {\rm crit}} \approx 8.7^{\circ}$ ($\alpha_{\rm mag, crit}\approx 0.3$) for highly magnetized thin disks (\citealt{Salvesen2016}) in prograde runs, it falls below this value for the retrograde run.

\section{Membrane paradigm}
\label{sect:mem_para}
\begin{figure*}
%\centering
\gridline{  \fig{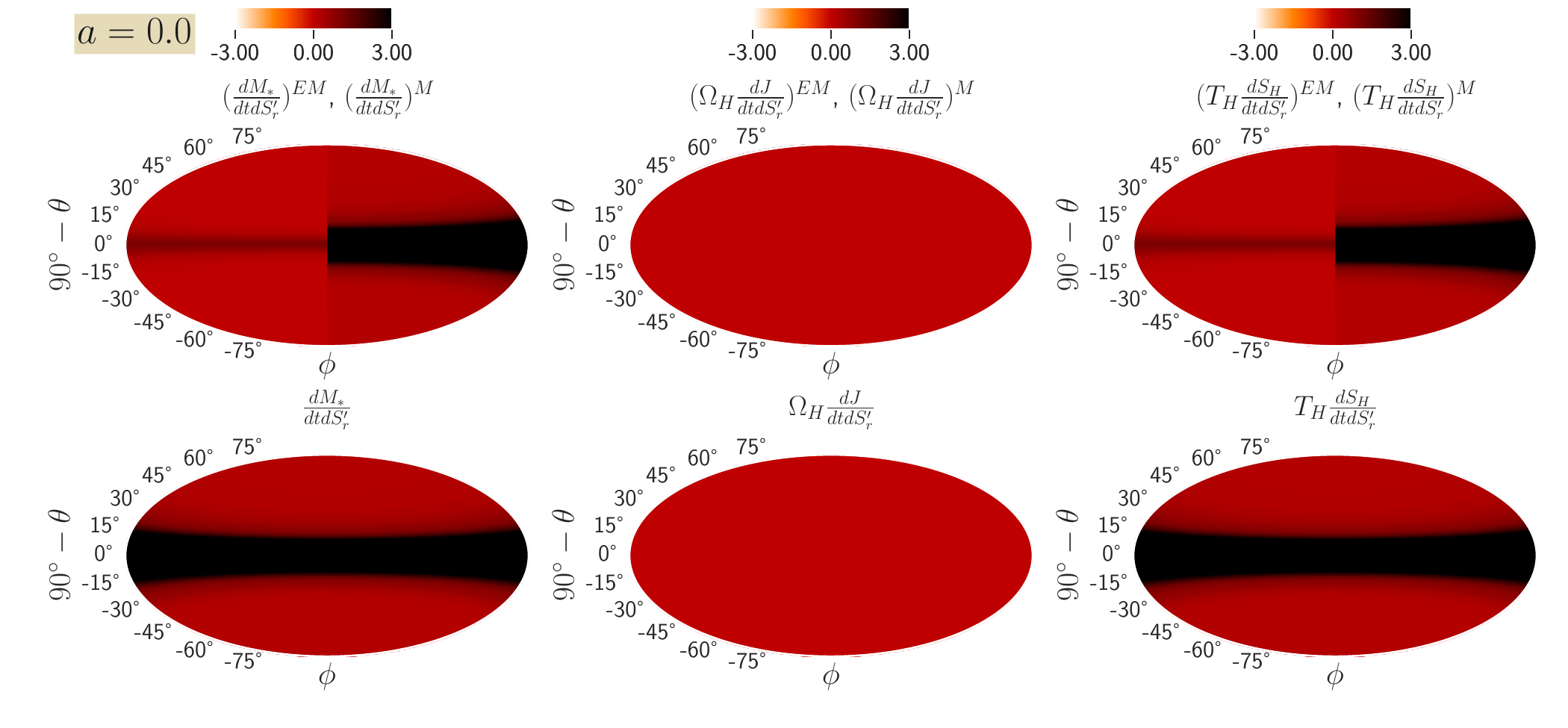}{0.95\textwidth}{(a)}        
          }
\gridline{
        \fig{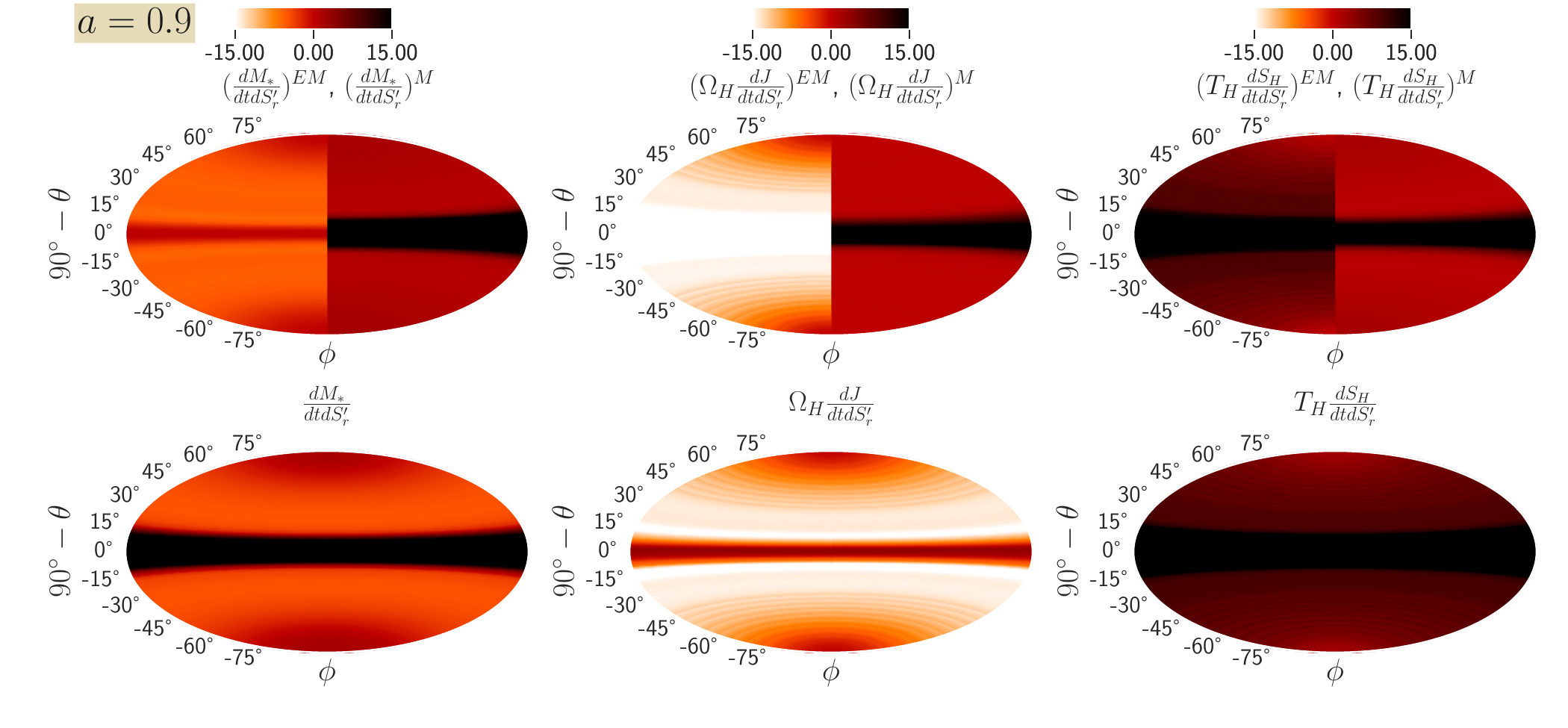}{0.95\textwidth}{(b)}
      }
\caption{(a) Distribution of different terms appear in the first law of black hole thermodynamics, describing mass-energy flux $dM/dtdS_r$, energy extracted by torque $\Omega_H dJ/dtdS_r$ and dissipation $T_HdS_H/dtdS_r$ on the membrane for the non-spinning BH in the quasi-stationary state. The top three panels compare the contributions from electromagnetic (EM) and matter components, with the left-hand side representing EM and the right-hand side representing matter. The bottom three panels show the combined EM and matter contributions to the mass-energy flux, energy extracted by torque, and dissipation, respectively.\\
(b) Same as Fig. \ref{fig:ent_torw_dis}(a), but for the spinning BH with spin $a=0.9$.}
\label{fig:ent_torw_dis}
\end{figure*}

The membrane paradigm (\citealt{Thorne1986_book}) has been extensively studied by \cite{Penna2013} using GRMHD simulations of geometrically thick hot flows. Here, we revisit some aspects of the membrane paradigm for a geometrically thin MAD as supplementary information to section \ref{sect:Bz_intro}.
We can recast equation (\ref{eq:first_law}), representing the first law of BH thermodynamics, as
\be
\label{eq:first_law_appen}
\frac{dM_{\ast}}{dt} = \Omega_H \frac{dJ}{dt} + T_H \frac{dS_H}{dt},
\ee 
where,
\bea 
\label{eq:dm_dt}
&& \frac{dM_{\ast}}{dt} = \int_{r_H} \alpha T^{\hat{r}}_{t_{BL}} \ dS^{\prime}_r \\
\label{eq:torque}
&& \Omega_H \frac{dJ}{dt} = - \Omega_H \int_{r_H} \alpha T^{\hat{r}}_{\phi_{BL}} \ dS^{\prime}_r \\
\label{eq:dissipation}
&& T_H \frac{dS_H}{dt} = \int_{r_H} \alpha^2 T^{\hat{r}}_{\hat{t}} \ dS^{\prime}_r
\eea 
representing mass energy entering/leaving the BH membrane, power generated by the torque applied on the membrane and dissipation occurring on the membrane, respectively. Here, we assume the following convention: if the BH membrane gains energy, $dM_{\ast}/dt > 0$, and vice versa; if rotational energy is extracted from the spinning BH, then $\Omega_H dJ/dt <0$, otherwise it is positive. 

Figures \ref{fig:ent_torw_dis}(a) and \ref{fig:ent_torw_dis}(b) compare the rates of mass-energy flux $dM/dtdS_r$, energy generated by torque $\Omega_H dJ/dtdS_r$ and dissipation $T_HdS_H/dtdS_r$ per unit area on the membrane for non-spinning ($a=0$) and spinning ($a=0.9$) BHs, respectively. For the non-spinning BH, the torque on the membrane vanishes, and all accreted mass-energy ($dM_{\ast}/dtdS_r>0$) is dissipated on the membrane. In contrast, for a spinning BH, the membrane experiences a negative torque ($\Omega_H dJ/dt <0$) due to magnetic fields, extracting rotational energy, while accreting matter imparts a positive torque ($\Omega_H dJ/dt >0$) primarily in the disk region, reducing the effectiveness of energy extraction. The extracted energy is partially converted into mass-energy flux into/out of the membrane, with the remainder being dissipated on the membrane. Notably, most of the mass-energy leaving the membrane at high latitudes, between the polar region and disk, is consistent with Fig.~\ref{fig:T_rt_em_lat}. Furthermore, dissipation is always positive, indicating that the BH entropy always increases, in accordance with the second law of BH thermodynamics.

\section{Comments of the calculation of radiative efficiency}
\label{sect:eta_appendix}
\begin{figure}
\centering
 \includegraphics[scale=0.6]{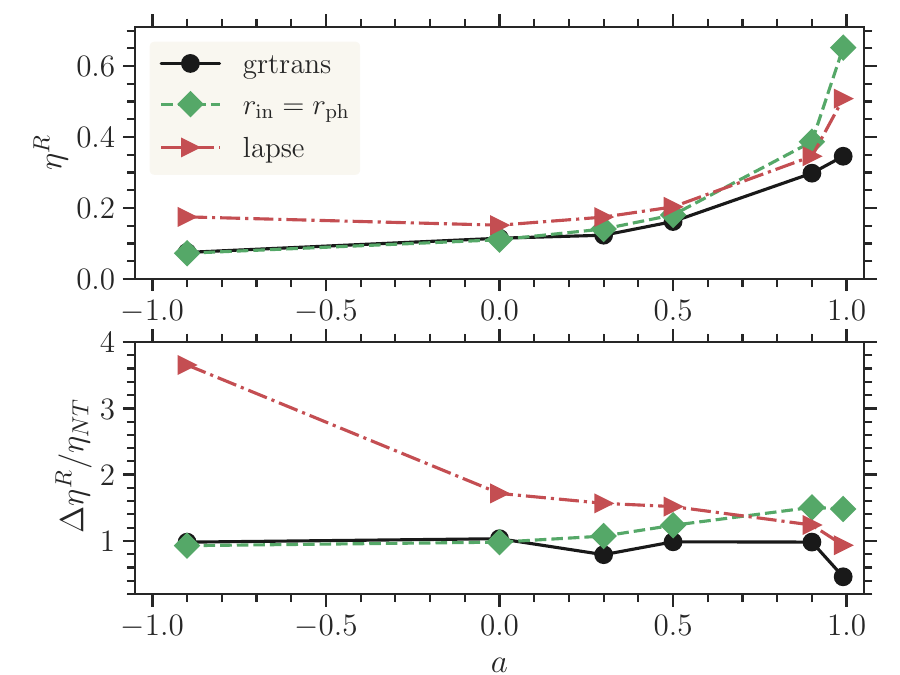}
\caption{Comparison of radiative efficiencies $\eta^{R}$ calculated using different methods. We recommend that ray-tracing is the preferred method for calculating radiative efficiency.}
\label{fig:eta_R_comp}
\end{figure}

We used ray tracing to calculate the radiative efficiency $\eta^{R}$ of the disk. However, it is also possible to calculate the radiative efficiency of the disk by integrating cooling term $\mathcal{S}_t$ over the four-volume $dtdV$ and  is given by
\begin{equation}
\label{eq:eta_rad}
\eta^{R}(r) = -\frac{1}{\dot{M}_H \left(t_f -t_i \right)} \int_{t_i}^{t_f} \int_{r_{\rm cap}}^{r} \int_{\theta=0}^{\theta=\pi} \int_{\phi=0}^{2 \pi} \mathcal{S}_t \ dt \ dV,
\end{equation}
where we account for the fact that not all emitted photons reach the observer at infinity due to capture by the BH (\citealt{Thorne1974}) or effects of redshift and beaming. The inner integration limit $r_{\rm cap}$ is often chosen to be the photon orbit radius $r_{\rm ph}$ (equation 2.18 of \citealt{Bardeen1972}). Alternatively, we can multiply the integral by the lapse function $\alpha_{\rm lapse}=1/\sqrt{-g^{tt}}$ and set $r_{\rm cap}=r_H$ to gradually absorb photons.

Figure \ref{fig:eta_R_comp} compares the radiative efficiency $\eta^{R}$
obtained using equation (\ref{eq:eta_rad}) with $r_{\rm cap}=r_{\rm ph}$ and the lapse function to the results obtained using ray-tracing. We find that neither the photon orbit radius nor the lapse function provides the ray-tracing values. Therefore, ray-tracing is the preferred method for calculating radiative efficiency.



\end{document}